\documentclass[fleqn,usenatbib]{mnras}

\usepackage{newtxtext,newtxmath}
\usepackage[T1]{fontenc}

\DeclareRobustCommand{\VAN}[3]{#2}
\let\VANthebibliography\thebibliography
\def\thebibliography{\DeclareRobustCommand{\VAN}[3]{##3}\VANthebibliography}

\usepackage{graphicx}	% Including figure files
\usepackage{amsmath}	% Advanced maths commands
\usepackage{multirow,booktabs}

\newcommand{\msun}{\mathrm{M}_{\sun}}
\newcommand{\zsun}{\mathrm{Z}_{\sun}}
\newcommand{\msunyr}{\msun\,\mathrm{yr}^{-1}}
\newcommand{\cc}{\mathrm{cm}^{-3}}
\newcommand{\dt}{\mathit{\Delta} t}
\newcommand{\dtt}{\mathit{\Delta} \tilde{t}}
\newcommand{\tff}{t_\mathrm{ff}}
\newcommand{\tffad}{t_\mathrm{ff,ad}}
\newcommand{\tffth}{t_\mathrm{ff,th}}
\newcommand{\geff}{\gamma_\mathrm{eff}}
\newcommand{\gad}{\gamma_\mathrm{ad}}
\newcommand{\Orot}{\Omega_\mathrm{rot}}
\newcommand{\Erot}{E_\mathrm{rot}}
\newcommand{\Egrav}{E_\mathrm{grav}}
\newcommand{\Eth}{E_\mathrm{th}}
\newcommand{\nad}{n_\mathrm{ad}}
\newcommand{\nth}{n_\mathrm{th}}
\newcommand{\nc}{n_\mathrm{0,c}}
\newcommand{\niso}{n_\mathrm{iso}}
\newcommand{\mc}{m_\mathrm{c}}
\newcommand{\mcmax}{m_\mathrm{c,max}}
\newcommand{\Nc}{N_\mathrm{c}}
\newcommand{\Ncb}{N_\mathrm{c,b}}
\newcommand{\Ncbmax}{N_\mathrm{c,b,max}}
\newcommand{\jeans}{\lambda_\mathrm{J}}
\newcommand{\jeansdx}{\jeans/\mathit{\Delta}x}

\title[Metal-poor disc fragmentation]{Gravitational Fragmentation of Extremely Metal-poor Circumstellar Discs}

\author[K. Shima \& T. Hosokawa]{
Kazuhiro Shima,$^{1,2}$\thanks{E-mail: kazuhiro\_shima@tap.scphys.kyoto-u.ac.jp  (KS)}
and Takashi Hosokawa$^{1}$\thanks{E-mail: hosokawa@tap.scphys.kyoto-u.ac.jp  (TH)}
\\
$^{1}$
Department of Physics, Graduate School of Science, Kyoto University, Sakyo, Kyoto 606-8502, Japan\\
$^{2}$
i-TEC Hankyu Hanshin Co., Ltd., Hanshin Noda Center Building 1-1-31 Ebie, Fukushima, Osaka 553-0001, Japan
}

\date{Accepted XXX. Received YYY; in original form ZZZ}

\pubyear{2021}

\begin{document}
\label{firstpage}
\pagerange{\pageref{firstpage}--\pageref{lastpage}}
\maketitle

% Abstract of the paper (!! word limit: 250 !!)
\begin{abstract}
We study the gravitational fragmentation of circumstellar discs accreting extremely metal-poor ($Z \leq 10^{-3}\,\zsun$) gas, performing a suite of three-dimensional hydrodynamic simulations using the adaptive mesh refinement code {\it Enzo}. We systematically follow the long-term evolution for $2 \times 10^3$ years after the first protostar's birth, for the cases of $Z=0$, $10^{-5}$, $10^{-4}$, and $10^{-3}\,\zsun$.
We show that evolution of number of self-gravitating clumps  qualitatively changes with $Z$. Vigorous fragmentation induced by dust cooling occurs in the metal-poor cases, temporarily providing $\sim 10$ self-gravitating clumps at $Z = 10^{-5}$ and $10^{-4}\,\zsun$. However, we also show that the fragmentation is a very sporadic process; after an early episode of the fragmentation, the number of clumps continuously decreases as they merge away in these cases. The vigorous fragmentation tends to occur later with the higher $Z$, reflecting that the dust-induced fragmentation is most efficient at the lower density. At $Z = 10^{-3}\,\zsun$, as a result, the clump number stays smallest until the disc fragmentation starts in a late stage. We also show that the clump mass distribution also depends on the metallicity. A single or binary clump substantially more massive than the others appear only at $Z = 10^{-3}\,\zsun$, whereas they are more evenly distributed in mass at the lower metallicities. We suggest that the disc fragmentation should provide the stellar multiple systems, but their properties drastically change with a tiny amount of metals.
\end{abstract}


%%%%%%%%%%%%%%%%%%%%%%%%%%%%%%%%%%%%%%%%%%%%%%%%%%
%%%%%%%%%%%%%%%%% BODY OF PAPER %%%%%%%%%%%%%%%%%%
%%%%%%%%%%%%%%%%%%%%%%%%%%%%%%%%%%%%%%%%%%%%%%%%%%

%%%%%%%%%%%%%%%%%%%%%%%
\section{Introduction}
\label{sec:intro}
%%%%%%%%%%%%%%%%%%%%%%%

The gravitational fragmentation of a circumstellar disc is a possible process that provides multiple stellar systems in the present-day Universe \citep[e.g.][]{Kratter16}. In fact, observations are revealing direct images of this process operating in some nearby star-forming regions \citep[][]{Tobin16,Ilee18}. Theoretical studies investigate necessary conditions for the disc fragmentation to yield self-gravitating clumps \citep[e.g.][]{Gammie01,Takahashi16}. Numerical simulations demonstrate that the vigorous fragmentation occurs particularly in an early evolutionary stage when the disc accretes the gas infalling from a surrounding envelope \citep[e.g.][]{Vorobyov2010,Machida11,Tsukamoto13,Oliva2020}.   

%----------------------------------------------------------%

Numerical simulations are a powerful tool to investigate the star formation in the early Universe, where only the pristine ($Z=0$) or extremely metal-poor (EMP, $Z \leq 10^{-3}\,\zsun$) gas exists \citep[see][for a review]{Greif15,Klessen19,Haemmerle20}. 
Despite significant differences from the present-day star formation, particularly in the gas thermal evolution, a circumstellar disc embedded in the accretion envelope commonly appears for this case \citep{Tan04,Yoshida08,Hirano14}. Although there is some diversity, three-dimensional (3D) simulations broadly show that the disc becomes gravitationally unstable and easily fragments for the primordial cases \citep{Saigo04,Machida08,Stacy10,Clark11,Smith11,Greif12,Vorobyov13,Susa13,Hosokawa16,Stacy16,Regan18,Sharda19,Sugimura20,Kimura20}.  

%-------------------------------------------------%

Despite the consensus that the disc fragmentation occurs in the primordial star formation, it is still challenging to predict the statistical properties of multiple stellar systems that finally appear \citep[e.g.][]{Stacy13, Sharda20}. This is due to limitations in the numerics used in the literature.
In the primordial star formation, the effective adiabatic index $\geff$ exceeds the critical value $4/3$ for the density $n \gtrsim 10^{20}\,\cc$, with which the hydrostatic protostellar structure appears\footnote{Note that $\geff$ also depends on the temperature, particularly when the gas is predominantly in the molecular state \citep[e.g.][]{Boley07,Sharda19}.} \citep{Omukai98,Yoshida08}. 
In most of previous simulations that study disc fragmentation, however, such dense gas is not spatially resolved to prevent the timestep from becoming very small. 
Some authors employ the so-called sink method, i.e., introduce a point mass which absorbs the nearby dense gas, representing an accreting protostar \citep[e.g.][]{Bate95, Krumholz04}. 
Another often-used method is assuming a stiff artificial equation of state (EOS) above a threshold density $\nth$ \citep[e.g.][]{Machida15, Hirano17}. Regardless of technical differences, these methods effectively mask the dense gas with $n \gtrsim \nth$, and the value of $\nth$ differs in different studies. There is a trend that simulations assuming the higher $\nth$ find the fragmentation in the earlier stage of the protostellar accretion \citep[e.g. see appendix in][]{Machida13}. Since increasing $\nth$ results in the short timestep, a higher-resolution simulation tends to only follow the shorter-term evolution to save the computational cost. The above partly explains why the number of fragments reported in the literature ranges from a few to $\sim 100$.

%-----------------------------------------------%

Recently, \citet{Susa19} has provided a comprehensive viewpoint in such a disputed situation. Considering the almost scale-free nature of the governing equations of the fluid dynamics with self-gravity, he shows that the apparently divergent results in the literature may suggest the same evolutionary trend described as
\begin{equation}
\Ncb \simeq 3 
\left( \frac{\tffad}{\tffth} \dt \right)^{0.3} \equiv 3\dtt^{0.3},
\label{eq:nfrag}
\end{equation}
where $\Ncb$ is the number of the gravitationally-bound clumps in a given snapshot, $\dt$ is the elapsed time since the first appearance of a protostar in the unit of year, and $\tffth$ and $\tffad$ are the free-fall timescales defined with the threshold number density $\nth$ and $\nad = 10^{19}\,\cc$, and $\dtt \equiv \sqrt{\nth / \nad} \dt$.
\citet{Susa19} shows that the simulation results by different authors roughly follow equation \eqref{eq:nfrag} once scaled with assumed values of $\nth$, though associated with an order-of-magnitude scatter.  

%-----------------------------------------------%

Although it is still unknown why the disc fragmentation for the primordial cases is well described by the simple relation such as equation \eqref{eq:nfrag}, an important fact is that a barotropic EOS with $\geff \simeq 1.1$ approximately represents the gas thermal evolution during the cloud collapse for $n \lesssim 10^{19}\,\cc$ \citep{Omukai98}. In fact, several simulations study the disc fragmentation assuming the same barotropic EOS with $\geff = 1.1$ for $n \leq \nth$, resulting in the evolution described by equation \eqref{eq:nfrag} \citep{Susa19}. Since $\nth$ is only the characteristic quantity for this case, the simple scaling of equation \eqref{eq:nfrag} may be convincing. 
This suggests that the disc fragmentation with a different EOS should provide different evolution of $\Ncb$. For instance, it is well known that adding a tiny amount of heavy elements and dust grains alters the EOS of a collapsing cloud \citep[e.g.][]{Omukai00, Bromm01,Omukai05, Schneider06, Smith08, Jappsen09, Schneider12, Safranek-Shrader14, Chiaki15, Chiaki16}. While previous studies demonstrate that the dust cooling enhances the fragmentation during the cloud collapse \citep[e.g.][]{Meece14,Smith15,Chiaki19}, its effect on the disk fragmentation remains to be further explored. 
\cite{Tanaka14} investigate the evolution of the circumstellar disc in metal-poor environments developing one-dimensional semi-analytic models. They predict that the discs with $Z \sim 10^{-5} - 10^{-3}\,\zsun$ are subject to the efficient dust cooling and are more unstable than those for the primordial cases.
\citet{Machida15} consider the disc fragmentation with various metallicities $0 \leq Z \leq 1\,\zsun$, performing a suite of 3D numerical simulations. 
They find qualitative differences between the cases with $Z \lesssim 10^{-4}\,\zsun$ and $Z \gtrsim 10^{-4}\,\zsun$; the vigorous disc fragmentation only occurs for the former metal-poor cases. Whereas \citet{Machida15} use the metallicity-dependent barotropic EOS, \citet{Chiaki20} recently report 3D simulations of the disc fragmentation solving the energy equation with relevant thermal processes coupled with a non-equilibrium chemical network. 
They find that for the metal-poor cases with $Z \leq 10^{-3}\,\zsun$, the disc fragmentation does not necessarily prevent the mass growth of the most massive protostar as many clumps are short-lived owing to the frequent merger or tidal disruption events. Although the above studies suggest the metallicity-dependence of the disc fragmentation, they both only follow the short-term evolution for $\sim 100\,\mathrm{yr}$ since the first emergence of a protostar.

%-------------------------------------------------------%

In this paper, we investigate the long-term evolution of the disc fragmentation with various metallicities in the range of $Z \leq 10^{-3}\,\zsun$: $Z=0$, $10^{-5}$, $10^{-4}$, and $10^{-3}\,\zsun$.
We systematically perform 3D hydrodynamic simulations that follow the evolution in the early collapse stage and subsequent accretion stage at the different metallicities. 
We examine the metallicity-dependence of the disc fragmentation that occurs during the first $2 \times 10^3\,\mathrm{yr}$ of the protostellar accretion. We track the evolution of the number of self-gravitating clumps for each case, as compiled for the primordial cases by \citet{Susa19}. We consider how the simple scaling given by equation \eqref{eq:nfrag} may be applicable or modified for the low-metallicity cases. We also investigate how other clump properties, such as their mass distribution, change with increasing the metallicity.

%--------------------------------------------------------%

The rest of the paper is organized as follows. We describe the numerical simulation methods in Section~\ref{sec:method}. We show our simulation results in Section~\ref{sec:result}. We finally provide discussion and concluding remarks in Sections~\ref{sec:discussion} and \ref{sec:conclusion}.

%%%%%%%%%%%%%%%%%%%%%%%%%%%%%%%%%%%%%%%%
\section{NUMERICAL SIMULATION METHODS}
\label{sec:method}
%%%%%%%%%%%%%%%%%%%%%%%%%%%%%%%%%%%%%%%%%

%-------------------------------------------------------%
\subsection{Equation of state at different metallicities}
\label{ssec:EOS}
%-------------------------------------------------------%

%---------------------- Fig.1 ------------------------%
\begin{figure}
  \includegraphics[width=\columnwidth]{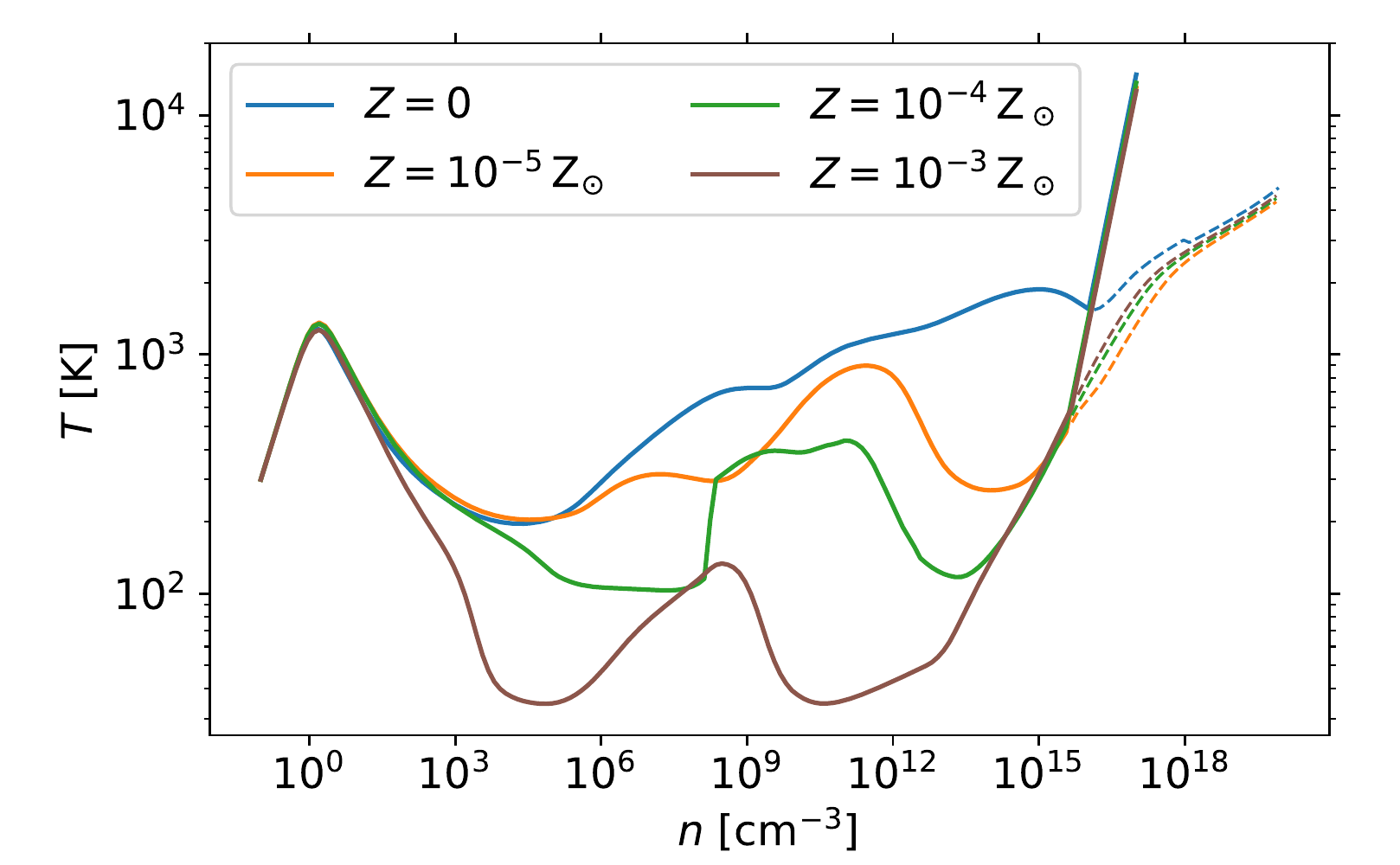}
  \caption{Barotropic equation of state (EOS) assumed for our simulations at different metallicities. The different colors represent different metallicities of $Z = 0$ (blue), $10^{-5}$ (orange), $10^{-4}$ (green), and $10^{-3}\,\zsun$ (brown). Variation of the temperature as functions of the number density is evaluated using the one-zone modeling of a collapsing cloud \citep[][]{Omukai05,Susa15}. The steep parts for $n \gtrsim 10^{16}\,\cc$ correspond to the stiff EOS assumed as the pressure floor (also see text). The dashed lines represent the original one-zone modeling results for which no pressure floor is imposed. 
  }
  \label{fig:n_vs_t}
\end{figure}
%---------------------------------------------------%

In this study, we model the gas thermal evolution with the pre-calculated barotropic EOS as in \citet{Machida15}, using the results of the so-called one-zone modeling of a collapsing cloud \citep[e.g.][]{Omukai00,Wakelam12,Grassi14,Smith17}. 
Fig.~\ref{fig:n_vs_t} shows the EOS we use, as variations of the temperature against the density at different metallicities of $Z=0$, $10^{-5}$, $10^{-4}$, and $10^{-3}\,\zsun$. The curve for $Z = 0$ is the same as that used in \citet{Susa15}, and those for the other cases are taken from \citet{Omukai05}. 
The one-zone models we consider assume that the cloud is in the free-fall collapse. The size of the collapsing homogeneous central core is approximated by the Jeans length. \citet{Omukai05} assume the dust-to-gas mass ratio linearly scaling with $Z$ and the standard Mathis-Rumpl-Nordsieck (MRN) dust size distribution in the solar neighborhood \citep{Mathis77}.
We do not consider cases with $0 < Z \leq 10^{-6}\,\zsun$ because the resulting EOS is almost identical to that for the primordial case. The present-day temperature of the cosmic microwave background (CMB) is assumed for these cases. 
\citet{Schneider10} study effects of enhancing the CMB temperature supposing the high redshifts up to $z \simeq 20$, and they find little impact for the EMP cases. The EOS curves presented in Fig.~\ref{fig:n_vs_t} do not perfectly match those provided by the {\it Enzo} source paper \citep[see Fig.~16 in][]{Enzo}, particularly for the low-metallicity cases. The discrepancy comes from the different implementations of metal cooling.

%------------------------------------------------------------------------------------%

In Fig.~\ref{fig:n_vs_t}, the local minima of the curves correspond to the points of $\geff = 1$, where the fragmentation is expected to be enhanced \citep[e.g.][]{Larson85,Larson05,Jappsen05}. The key concept here is that a self-gravitating cloud easily evolves into the filamentary configuration, and such a filament continues to collapse as far as $\geff < 1$. When $\geff > 1$, the filament ceases to collapse and breaks up into fragments.
We see that the curves except for $Z=0$ have the double minima below and above $n \sim 10^9\,\cc$. The former is caused by molecular cooling and the latter by cooling via dust thermal emission. Since $n \gtrsim 10^9\,\cc$ in circumstellar discs we consider, only the dust-induced fragmentation is relevant to our cases.

%-----------------------------------------------------------------%

Although the barotropic EOS is obtained for the evolution during the early collapse stage, it has been applied to study the disc fragmentation in the literature. \citet{Clark11} show that, at least for the primordial cases, applying the barotropic EOS tends to result in the lower disc temperature and thus more fragmentation than solving the thermal and chemical processes in a time-dependent hydrodynamic code \citep[see also][]{Matsukoba21}. Nonetheless, \citet{Susa19} has found that the previous simulations with both approaches show similar evolution of the clump number as described by equation \eqref{eq:nfrag}. Our current method is thus valid for the aim of the current work, whereas quantitative effects of relying on the barotropic EOS is to be examined in future studies.

%-------------------- TABLE 1 ----------------------%
\begin{table}
  \centering
  \caption{Cloud initial parameters: radius, mass, temperature, and average density.}
  \label{tab:cloud_params}
  \begin{tabular}{lcccc}
    \hline
    Profile  & $r_0$                 & $M_0$                 & $T_0$ & $\bar{n}_0$ \\
             & (pc)                  & ($\msun$)             & (K)   & ($\cc$)     \\
    \hline
    cloud A & 1.3                    & 1.1 $\times$ 10$^{3}$ & 195   & 3.8 $\times$ 10$^{3}$ \\
    cloud B & 2.5 $\times$ 10$^{-1}$ & 3.6 $\times$ 10$^{1}$ & 35    & 1.9 $\times$ 10$^{4}$ \\
    \hline
  \end{tabular}
\end{table}
%----------------------------------------------------%
%--------------------- TABLE 2 ----------------------%
\begin{table}
  \centering
  \caption{Cases considered.}
  \label{tab:simulation_params}
  \begin{tabular}{ccc}
    \hline
    Z         & initial profile & $\beta$          \\
    ($\zsun$) &                 &                  \\
    \hline
    0         & cloud A         & 0.03, 0.06, 0.09 \\
    10$^{-5}$ & cloud A         & 0.03, 0.06, 0.09 \\
    10$^{-4}$ & cloud A         & 0.03, 0.06, 0.09 \\
    10$^{-3}$ & cloud B         & 0.03, 0.06, 0.09 \\
    \hline
  \end{tabular}
\end{table}
%---------------------------------------------------%

%------------------------------%
\subsection{Simulation setup}
\label{ssec:setup}
%------------------------------%

Our simulations use the adaptive mesh refinement (AMR) hydrodynamics code, {\it Enzo} \citep{Enzo}. The gas is evolved with self-gravity solving the Poisson equation using a multi-grid solver. Regarding the solver of the hydrodynamics, we adopt a three-dimensional implementation of the {\it Zeus} hydro-code \citep{Zeus1, Zeus2}.

%-----------------------------------------------------%

As for the initial states, we assume two different cloud properties: cloud A for cases with $Z \leq 10^{-4}\,\zsun$ and cloud B with $10^{-3}\,\zsun$ (see Tables \ref{tab:cloud_params} and \ref{tab:simulation_params}), both of which rigidly rotate at the angular frequency $\Orot$.
The cloud takes the density profile of a Bonnor-Ebert (BE) sphere \citep{Ebert55,Bonnor1956}; a hydrostatic isothermal self-gravitating sphere of gas that is confined by its external pressure. While such a profile is derived analytically, cosmological simulations show that it is realized in a primordial star-forming cloud in the so-called "loitering" phase \citep{Bromm02,Hirano14}.
The BE profile is parametrized by the central density $\nc$ and temperature $T_0$, for which we use the values at the loitering point or the local minimum at $n \sim 10^4\,\cc$ in Fig.~\ref{fig:n_vs_t}. We use $\nc = 2.0 \times 10^{4}\,\cc$ and $T_0 = 195\,\mathrm{K}$ for the primordial case (cloud A). 
We assume that the critical BE profile continues until the cloud radius $r_0$, where the enclosed mass is $\simeq 10^3\msun$, a typical value for the primordial clouds \citep[][]{ABN02,Yoshida03,Stacy13,Hirano14}. 
We also assume the cloud is embedded in the homogeneous medium that provides the constant external pressure for $r > r_0$. We enhance the density by $10\,\%$ within the cloud to cause the collapse, and we further add a perturbation of $m=2$ mode as $\rho_0(r) (1 + \delta \cos{2\phi})$, where $\rho_0$ is the unperturbed density, $\delta$ the perturbation amplitude, and $\phi$ the azimuthal angle around the rotation (or $z$-) axis. We only consider the cases with $\delta = 0.1$ in the current work. The above setup of the initial perturbations is the same as in \cite{Susa19}. \cite{Machida15} also use the same $m=2$ mode perturbations, and they further add $m=3$ mode with the small amplitude of $\delta = 0.01$.  
We fix the initial perturbations in our simulations to make comparisons to the previous studies simple.

%------------------------------------------------------%

Fig.~\ref{fig:n_vs_t} shows that the curves for $Z \leq 10^{-4}\,\zsun$ are almost identical to the primordial case for $n \lesssim 10^4\,\cc$. We thus use the initial condition of the cloud A also for the low-metallicity cases with $Z \leq 10^{-4}\,\zsun$.
We only consider a different initial cloud configuration for $Z = 10^{-3}\,\zsun$, with which the evolution departs from the other cases at $n \sim 100\,\cc$. 
To characterize the BE sphere for this case, we use $\nc = 1.0 \times 10^{5}\,\cc$ and $T_0 = 35\,\mathrm{K}$, the values at the local minimum (cloud B). The following procedure to construct the initial state is the same as for cloud A. 

%-----------------------------------------------------%

We assume idealized clouds with artificial density perturbations as the initial conditions, and we also ignore the magnetic fields and turbulence for simplicity. 
As discussed separately in Section \ref{ssec:caveat}, these effects make the disc fragmentation more stochastic, which may obscure the metallicity dependencies we consider. We remark that our simulations follow the evolution of the protostellar accretion for $2 \times 10^3\,\mathrm{yr}$ (see Section~\ref{ssec:cases}), and the duration corresponds to the free-fall timescale at $n \sim 10^9\,\cc$. This is much higher than the initial central value, meaning that the disc only accretes the gas coming from a dense accretion envelope set during the cloud collapse.  
In our simulations, the mass of the most massive protostar remains smaller than $20~\msun$ except for the primordial cases. We ignore the protostellar radiative feedback, the impacts of which are expected to be limited during the simulation duration (see also Section \ref{ssec:caveat}). 
We also note that the disc fragmentation is not only a process that provides numerous self-gravitating clumps. Filamentary structure of star-forming clouds is known to cause fragmentation in general, particularly for the low-metallicity cases where the dust cooling operates \citep[e.g.][]{Tsuribe06,Clark08,Dopcke11,Dopcke13,Chiaki16,Sugimura17}. Our simple setup only allows the filamentary structure to develop in the circumstellar discs. We isolate the effects of the disc fragmentation with different metallicities in the current work.

%-----------------------------------------------------%

The simulation box size is 3\,pc for cloud A and 0.75\,pc for cloud B on a side, which are $\simeq 3$ times larger than the cloud radius $r_0$. The box is covered by 128 root grids initially. 
Higher grid levels are added during the evolution by the adaptive mesh refinement technique, reducing the cell size by a factor of two. Spatial cells are refined based on the requirement that the Jeans length must not fall below 32 cells \citep[e.g.][]{Federrath11}. We allow the de-refinement by doubling the cell size if a fine grid level is no longer necessary. We also confirm the numerical convergence of our results by performing additional simulations resolving the Jeans length by 16 and 64 cells (see Section~\ref{ssec:res}).
At the maximum refinement level 16 for cloud A and 14 for cloud B (the minimum cell size is the same at 0.074\,au
in both clouds), where the Jeans criteria inevitably must break, we introduce a pressure floor to halt the collapse at a finite density, preventing individual cells from becoming unphysically massive.
We realize the pressure floor by assuming the stiff EOS with $\gad = 2$ for $n \gtrsim 10^{16}\,\cc$ \citep{Takahira14}, as illustrated in Fig.~\ref{fig:n_vs_t}. The floor density $10^{16}\,\cc$ is similar to that used in \citet{Susa19}, who has only considered the primordial case and shown the evolution well described by equation \eqref{eq:nfrag}. We tested the effects of lowering the floor density to $10^{15}\,\cc$, by performing an experimental simulation for $Z=10^{-5}~\zsun$ with the default choice of the cloud's rotation (see Section~\ref{ssec:cases}). We did not find significant differences in the evolution of the number of self-gravitating clumps formed through disk fragmentation.

%------------------------------%
\subsection{Cases considered}
\label{ssec:cases}
%-----------------------------%

%------------------------------------------------------------------%

We calculate the evolution of 12 models with four different metallicities and three different degrees of the initial rotation. These models, as summarized in Table~\ref{tab:simulation_params}, allow us to study how much varying the cloud rotation affects the disk fragmentation compared to varying the metallicity. We represent the cloud's rotational degree with $\beta$, or the ratio of the rotation energy $\Erot$ to the gravitational energy $\Egrav$
\begin{equation}
\beta \equiv \frac{\Erot}{\left|\Egrav\right|} = \frac{\Orot^2 r_0^3}{3GM_0} .
\end{equation} 
We take $\beta = 0.06$ as a standard value, which corresponds to the initial angular frequency $\Orot \simeq 2.0 \times 10^{-14}\,\mathrm{s}^{-1}$ for cloud A and $\Orot \simeq 4.4 \times10^{-14}\,\mathrm{s}^{-1}$ for cloud B. We also consider the cases with $\beta = 0.03$ and 0.09 for comparisons.
Whereas our examined range of $\beta$ is almost the same as in \citet[][]{Susa19}, it is much higher than that in \citet{Machida15} by more than an order of magnitude. Cosmological simulations show that primordial clouds typically has $\beta \sim 0.1$ \citep{ABN02,Hirano14,Stacy10}, and they correspond to our cases with the most rapid rotation. 
The ratio of the cloud's thermal energy $\Eth$ to gravitational energy, often represented by $\alpha$ parameter \citep[e.g.][]{Miyama84}, 
\begin{equation}
\alpha \equiv \frac{\Eth}{\left|\Egrav\right|} 
= \frac{5 c_s^2 r_0}{2 G M_0} ,
\end{equation}
is $\simeq 0.93$ in our initial configuration of both clouds A and B, where $c_s$ is the sound speed. We fix $\alpha$ for our cases following \cite{Machida15} and \cite{Susa19}. Since this parameter is directly related to mass supply rates onto a disc from the surrounding envelope, it is also a key parameter of disk fragmentation. It is thus important to systematically study such additional effects in future studies.

%-----------------------------------------------------------------%

We first follow the evolution of the early run-away collapse for all the cases we consider. The collapse continues until the first self-gravitating clump (or "protostar") appears when the refinement reaches the maximum level. After that, we follow the evolution of the protostellar accretion for $2 \times 10^3\,\mathrm{yr}$, during which the disc fragmentation occurs. The duration is comparable to that in \citet{Susa19}, but it is about ten times longer than in \citet{Machida15} and \citet{Chiaki20}. Whereas \citet{Susa19} only considers the primordial cases, we study the metallicity-dependence of the disc fragmentation during the similar long-term evolution in 3D.

%%%%%%%%%%%%%%%%%%%%%%%%%%%%%%%%%%%%%%%%%%%%%%%%%%%%%
%--------------------- Fig.2 -----------------------%
\begin{figure*}
  \includegraphics[width=1.9\columnwidth]{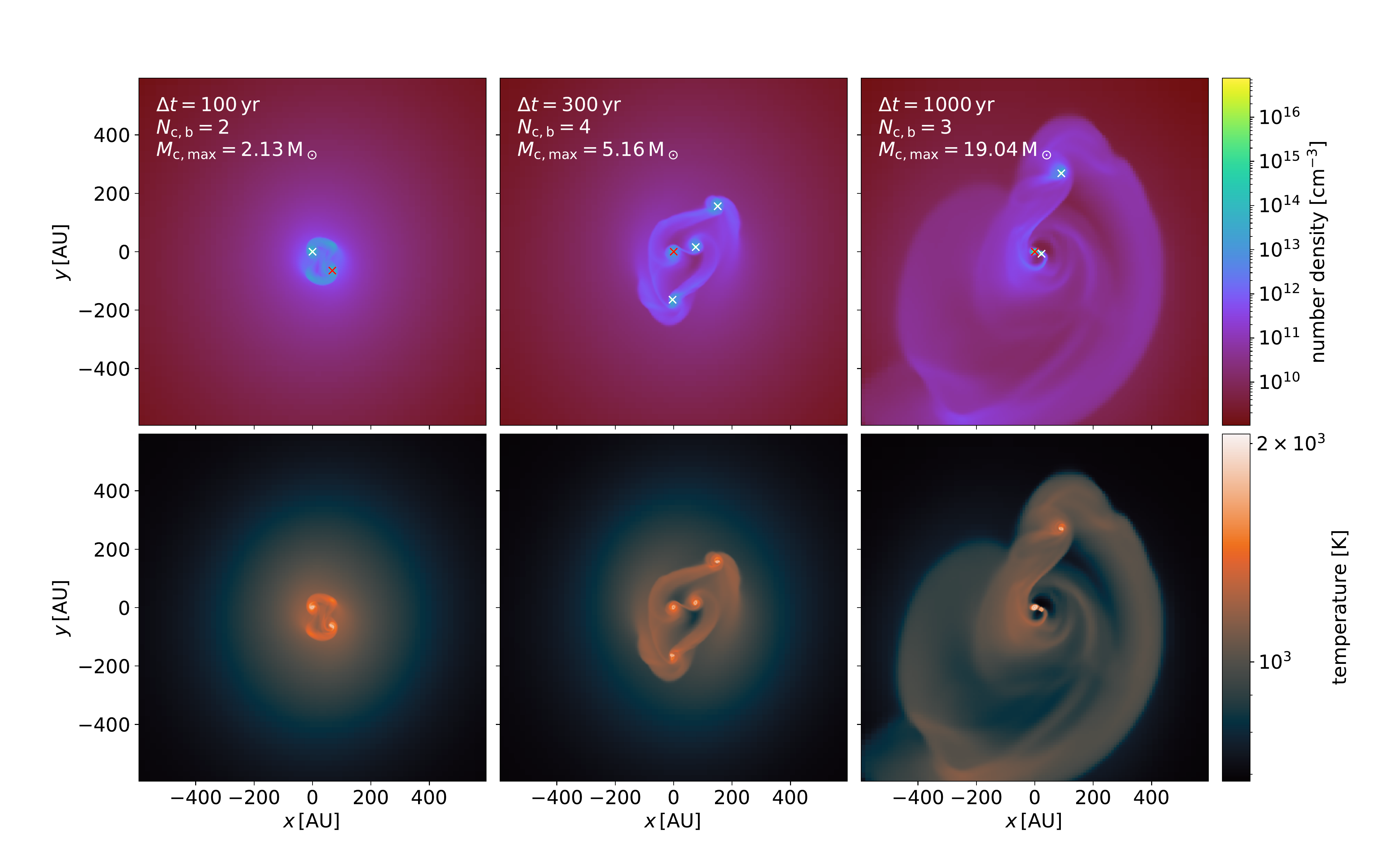}
  \caption{Images of the density-weighted projection maps (through the z-axis) of the number density (top) and temperature (bottom) for the case of $Z = 0$ and $\beta = 0.06$. The left, middle, and right panels show the snapshots at $\dt = 100\,\mathrm{yr}$, $300\,\mathrm{yr}$, and $10^3\,\mathrm{yr}$ after the first appearance of a gravitationally bound clump. We use the color scales covering the maximum and minimum values that appear through the snapshots for both the top and bottom panels. The center of each panel corresponds to the position of the densest cell. The white crosses in the upper panels indicate the mass-center positions of the gravitationally bound clumps identified by the finder with a threshold of $10^{14}\,\cc$ (see Section~\ref{sec:clump_finder}). The red cross denotes the most massive clump at each snapshot. 
  The disc radially spreads from left to right as it accretes the gas infalling from the envelope with the higher angular momentum.}
  \label{fig:Z0b6_projection}
\end{figure*}
%------------------------------------------------%
%--------------------- TABLE 3 ----------------------%
\begin{table*}
  \centering
  \caption{Properties of the disc, arms, and self-gravitating clumps at different epochs}
  \label{tab:discs}
  \begin{tabular}{lcccccccccccc}
  \toprule
%%    & & disc & & & & arms &  & & clumps &  \\
\multicolumn{2}{c}{} & \multicolumn{5}{c}{disc} & \multicolumn{3}{c}{arms}  & \multicolumn{3}{c}{clumps} \\
    \cmidrule(lr){3-7} \cmidrule(lr){8-10} \cmidrule(lr){11-13} 
    Z         & $\dt$     & $n_\mathrm{disc,min}$ & $M_\mathrm{disc}$  & $\bar{R}_\mathrm{disc}$ & $\bar{n}_\mathrm{disc}$  & $\bar{T}_\mathrm{disc}$  & $M_\mathrm{arm}$ &  $\bar{n}_\mathrm{arm}$ & $\bar{T}_\mathrm{arm}$ & $M_\mathrm{c, max}$ & $M_\mathrm{c, tot}$ & $N_\mathrm{c,b}$ \\
    ($\zsun$) & (100 years)   & ($\cc$) & ($\msun$) & (AU) & ($\cc$)  & (K) & ($\msun$) & ($\cc$) & (K) & ($\msun$) & ($\msun$) \\
    \midrule
    0         & 1   & $5.57 \times 10^{11}$ & 11.0  & 88.4 & $5.36 \times 10^{12}$ & 1320  & 8.05 & $2.28 \times 10^{13}$ & 1490 & 2.13 & 4.15 & 2 \\
              & 3   & $2.15 \times 10^{10}$ & 28.7  & 296 & $1.95 \times 10^{11}$ & 1120 & 22.7  & $1.75 \times 10^{12}$ & 1240 & 5.16 & 14.8 & 4 \\
              & 10  & $4.92 \times 10^9$ & 60.7 & 583 & $6.64 \times 10^{10}$  & 1030 & 48.6 & $2.91 \times 10^{11}$ & 1150 & 19.0 & 32.4 & 3 \\
              & 20 & $3.90 \times 10^9$  & 92.8 & 886 & $4.80 \times 10^{10}$ & 996 & 71.5 & $5.40 \times 10^{11}$ & 1180 & 43.3 & 58.9 & 3 \\
\hline
   $10^{-5}$ & 1 & $1.28 \times 10^{11}$ & 11.0 & 170 & $4.74 \times 10^{11}$ & 885 & 4.69 & $1.37 \times 10^{13}$ & 327 & 1.26 & 3.34 & 13 \\
            & 3 & $1.44 \times 10^{10}$ & 20.2 & 305 & $1.32 \times 10^{11}$ & 876 & 14.2 & $6.92 \times 10^{11}$ & 858 & 3.22 & 8.20 & 8 \\
            & 10 & $5.09 \times 10^9$ & 35.2 & 493 & $6.04 \times 10^{10}$ & 812 & 25.5 & $3.12 \times 10^{11}$ & 897 & 14.3 & 14.4 & 3 \\
            & 20 & $2.14 \times 10^9$ & 52.6 & 720 &  $3.30 \times 10^{10}$ & 744 & 43.0 & $1.17 \times 10^{11}$ & 868 & 19.5 & 23.0 & 3 \\
            \hline
    $10^{-4}$ & 1 & $1.93 \times 10^{11}$ & 2.19 & 71.1 & $1.14 \times 10^{11}$ & 218 & 1.52 & $3.40 \times 10^{13}$ & 120 & 0.664 & 1.16 & 3 \\
            & 3 & $3.74 \times 10^{10}$ & 5.87 & 170 & $2.55 \times 10^{11}$ & 397 & 4.15 & $7.59 \times 10^{12}$ & 126 & 1.16 & 2.95 & 13 \\ 
            & 10 & $5.63 \times 10^9$ & 14.6 & 370 & $5.99 \times 10^{10}$ & 422 & 10.9 & $6.30 \times 10^{11}$ & 286 & 3.86 & 8.90 & 10 \\
            & 20 & $1.41 \times 10^9$ & 23.1 & 725 & $1.36 \times 10^{10}$ & 390 & 15.2 & $1.88 \times 10^{11}$ & 420 & 7.14 & 13.2 & 3 \\
            \hline
    $10^{-3}$ & 1 & $1.98 \times 10^{11}$ & 0.390 & 41.4 & $2.12 \times 10^{12}$ &  46.1 & 0.305 & $1.93 \times 10^{13}$ & 69.9 & 0.206 & 0.206 & 1 \\
        & 3 & $9.58 \times 10^{10}$ & 0.818 & 52.3 & $2.40 \times 10^{12}$ & 46.7 & 0.742 & $1.59 \times 10^{13}$ & 64.9 & 0.220 & 0.220 & 1 \\
        & 10 & $2.34 \times 10^{10}$ & 2.52 & 77.5 & $1.57 \times 10^{12}$ & 44.8 & 2.43 & $1.12 \times 10^{13}$ & 58.0 & 1.68 & 2.14 & 3 \\
        & 20 & $8.68 \times 10^9$ & 4.89 & 107 & $7.82 \times 10^{11}$ & 42.0 & 4.78 & $5.45 \times 10^{12}$ & 50.6 & 3.17 & 4.31 & 4 \\
    \bottomrule
  \end{tabular}
\end{table*}
%---------------------------------------------------%
%%%%%%%%%%%%%%%%%%%%%%%%%%%%%%%%%%%%%%%%%%%%%%%%%%%%%

%--------------------------%
\subsection{Clump finder}
\label{sec:clump_finder}
%--------------------------%

As described in Section~\ref{ssec:setup}, we artificially halt the cloud collapse by using the stiff EOS for $n \gtrsim
10^{16}\,\cc$. Otherwise, the collapse further continues, and the timestep becomes smaller and smaller. Following the long-term evolution of the disc fragmentation becomes computationally infeasible for such a case. An alternative method to achieve the same purpose is using the Lagrangian sub-grid model such as the sink cell/particle method. However, we do not resort to this in the current work. 

%-----------------------------------------------------------%

A disadvantage of the sink method is that the results may depend on the details of the implementation. For instance, 
different implementation assumes different criteria for the sink creation, accretion onto the sink, and mergers between them, though some possible solutions have been developed \citep[e.g.][]{Federrath10, Hubber13}. Since we also  implemented the sink method proposed by \citet{Federrath10} to {\it Enzo} \citep{Shima18}, we actually performed preliminary simulations of the disc fragmentation using our modified version of the code.  
However, it turned out that the simulation with the sink method is more computationally expensive than that with the stiff EOS.
We thus adopt the current method of the stiff EOS, which is simpler and more efficient than the sink method for our specific cases. 
\citet{Susa19} has investigated the effects of using the stiff EOS and sink methods in comparisons in smoothed particle hydrodynamics (SPH) simulations. Fortunately, equation \eqref{eq:nfrag} well describes the evolution of the disc fragmentation observed in both cases.

%-----------------------------------------------------------------%

Since our method defines neither self-gravitating clumps nor protostars, we need a method to detect them. We save the simulation data every 10 yr to follow the evolution of self-gravitating clumps after the pressure reaches the floor value. To identify the clumps at each snapshot, we use the finder implemented in {\it yt} \citep{Smith09, yt}, 
which enables detecting individual clumps that are disconnected from each other. We begin with the density threshold of $\niso = 10^{14}\,\cc$ to find iso-density contours and identify clump candidates. We continually multiply $\niso$ by a factor of 10 and apply the clump finding algorithm again. If a clump identified with the lower-density contour turns out to contain two clumps with the higher-density contour, we regard the number of clumps as two. We repeat the whole procedure until $\niso$ exceeds the maximum density in the given snapshot. We estimate the mass of a clump by summing up the gas contained in a disconnected iso-density contour at the lowest level, above which there are no further sub-clumps inside.

%-------------------------------------------------------------%

For each clump, we check whether it is gravitationally bound or not as follows. We consider $W + K + U < 0$ as the condition for the gravitational binding, where $W~(<0)$ is the total gravitational energy, $K$ the total kinetic energy, and $U$ the total thermal energy within the clump. We evaluate $W$ by summing up the gravitational binding energy between all two point cells inside the clump,
\begin{equation}
    W = - \sum_{i,j, i\neq j} \frac{G m_i m_j}{|\mathbfit{r}_i - \mathbfit{r}_j|} ,
\end{equation}
where $m_i$ is each cell mass and the $r_i$ is the position of the cell. The total kinetic energy is 
\begin{equation}
    K = \sum_{i} \frac{1}{2} m_i ((u_{i} - u_{c})^2 + (v_{i} - v_{c})^2 + (w_{i} - {w_{c}})^2)  ,
\end{equation}
where ($u_{i}$, $v_{i}$, $w_{i}$) is the velocity in each cell inside the clump, and ($u_{c}$, $v_{c}$, $w_{c}$) is the velocity of the clump's center-of-mass. We normally only count the gravitationally bound objects, but we also investigate how much the clump number increases if we skip the binding check in Section~\ref{ssec:res}. 
Note that we usually find $10^{3-5}$ cells within each self-gravitating clump. Since we do not impose a minimum number of cells, however, we count tiny structures with $\sim$ ten cells without the gravitational binding check. 
Hereafter, we define the origin of the elapsed time $\dt = 0$ as the epoch when the clump finder first detects a self-gravitating clump. 

%--------------------------------------------------%

We note that the initial contouring density $\niso = 10^{14}\,\cc$ roughly corresponds to the values where the EOS curves for $Z = 10^{-5}$ and $10^{-4}\,\zsun$ take the local minima owing to the dust cooling (see Fig.~\ref{fig:n_vs_t}).
We have tested different initial choices of $\niso = 10^{13}\,\cc$ and $10^{15}\,\cc$. The iso-density contours with $\niso = 10^{13}\,\cc$ cover a large part of the disc rather than the individual clumps. Starting the analysis with $\niso = 10^{15}\,\cc$, on the other hand, returns almost the same result as with $\niso = 10^{14}\,\cc$. 
We have also tested a different $\niso$-multiplying factor 2 instead of 10, with which the results hardly change.

%----------------------------------------------%
\subsection{Assessment of discs and arms}
\label{sec:discarm_finder}
%----------------------------------------------%

We further evaluate physical properties of discs and arms, which give birth to the self-gravitating clumps, for representative cases with the different metallicities. We first identify the disc for a given snapshot of simulation data. Since our simulation starts with the idealized initial condition as described in Section ~\ref{ssec:setup}, only the disc have the complex substructure such as arms. 
Making use of this fact, we look for the maximum density threshold below which the corresponding density contour is topologically connected.
The density contour splits into two or more parts above the threshold as we delineate the substructure within the disc. We regard this threshold as the minimum disc density,  $n_\mathrm{disc,min}$. We determine $n_\mathrm{disc,min}$ by the bisection method, repeating the contour drawing with different threshold densities. Once fixing $n_\mathrm{disc,min}$, we derive the disc mass $M_\mathrm{disc}$ by calculating the enclosed mass. 
By dividing $M_\mathrm{disc}$ by the volume, 
we obtain the disc's mean density $\bar{n}_\mathrm{disc}$, which is also converted to the mean temperature $\bar{T}_\mathrm{disc}$ by the EOS given in Fig.~\ref{fig:n_vs_t}. We evaluate the disc radius  $\bar{R}_\mathrm{disc}$ by averaging the distances between the most massive clump and disc's outer edge on the equatorial plane. 

%-------------------------------------------------------------%

As for the arms, we count the dense parts where the local density exceeds $10 \times n_\mathrm{disc,min}$. We calculate the arms' mass $M_\mathrm{arm}$, mean density $\bar{n}_\mathrm{arm}$, and mean temperature $\bar{T}_\mathrm{arm}$ in the same manner as for the disc. We also check whether the above procedures adequately capture the whole disc and arms by eye for every snapshot data we analyze. Table 3 below summarizes the results, which we discuss in Section \ref{ssec:dfrag}.

%%%%%%%%%%%%%%%%%%%%%%%%%%%%%%%%%%%%%%%%%%%%%%%%%%%%%
%------------------- Fig.3 ----------------------%
\begin{figure}
  \includegraphics[width=\columnwidth]{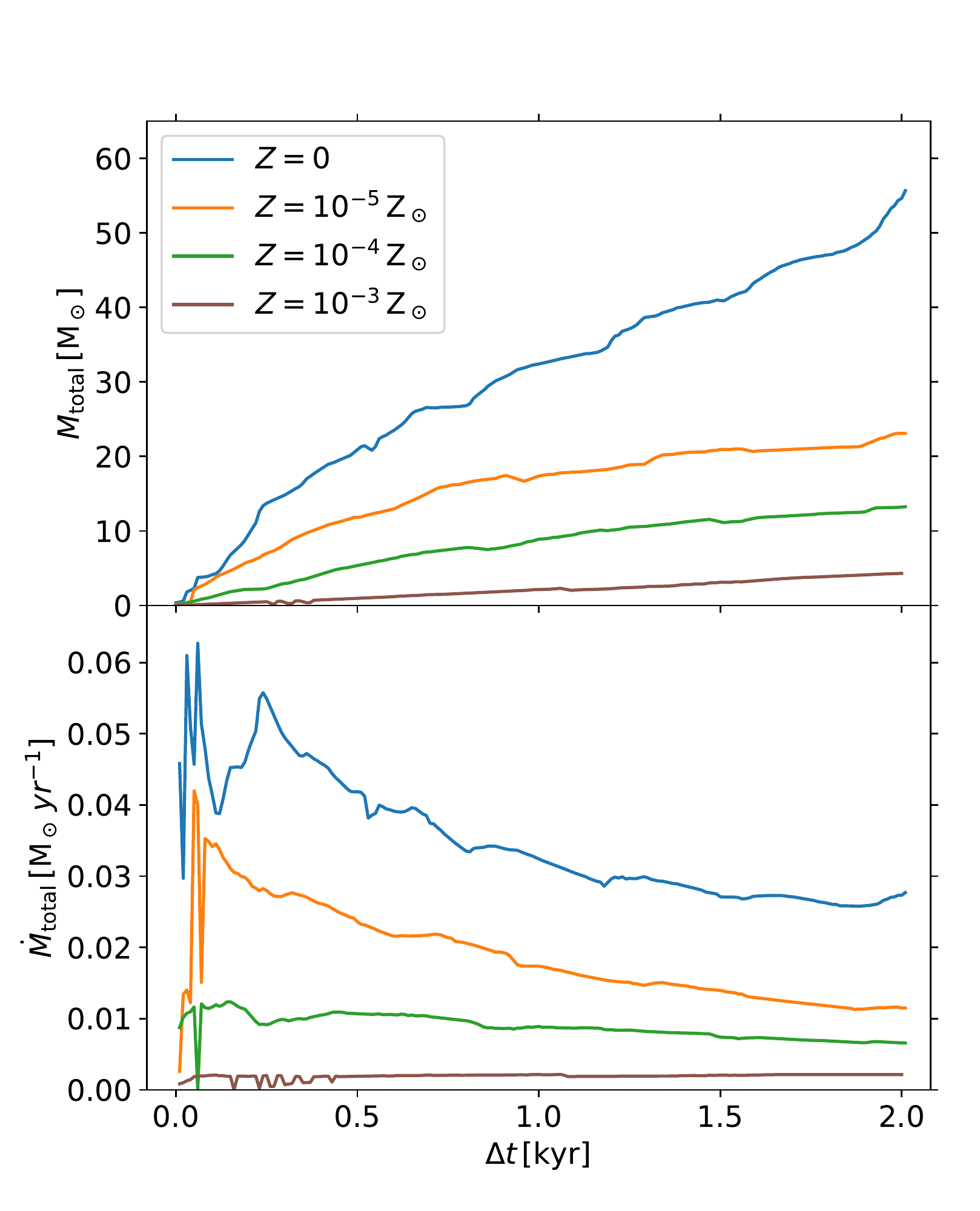}
  \caption{
  Time evolution of the total mass of self-gravitating clumps detected by the finder with the initial iso-contour density $\niso = 10^{14}\,\cc$ 
  (upper panel), and total accretion rates onto them (lower panel). 
  The line colors represent the same cases with different metallicities as in Fig.~\ref{fig:b6_susaplot}.
  }
  \label{fig:b6_mass}
\end{figure}
%-------------------------------------------------%
%--------------------- Fig.4 ---------------------%
\begin{figure*}
  \includegraphics[width=1.9\columnwidth]{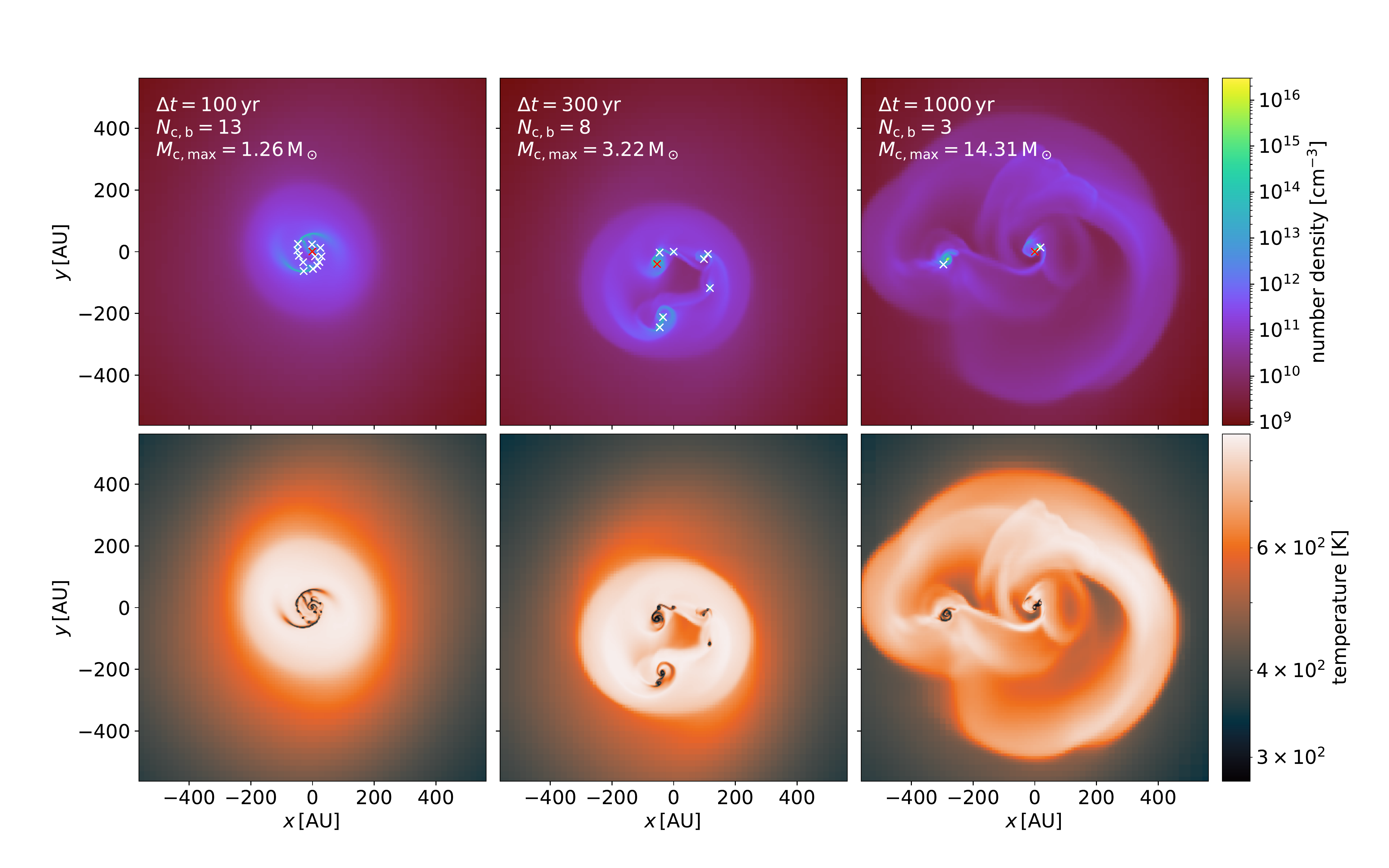}
  \caption{Same as Fig.~\ref{fig:Z0b6_projection} but for the case of $Z = 10^{-5}\,\zsun$. There are more than 10 clumps at the first snapshot $\dt = 100\,\mathrm{yr}$, but the clump number monotonically decreases until the last snapshot at $\dt = 10^3\,\mathrm{yr}$. Note that the color scaling for both the density and temperature is different from Fig.~\ref{fig:Z0b6_projection}. }
  \label{fig:Z5b6_projection}
\end{figure*}
%-----------------------------------------------%
%------------------- Fig.5 ---------------------%
\begin{figure*}
  \includegraphics[width=1.9\columnwidth]{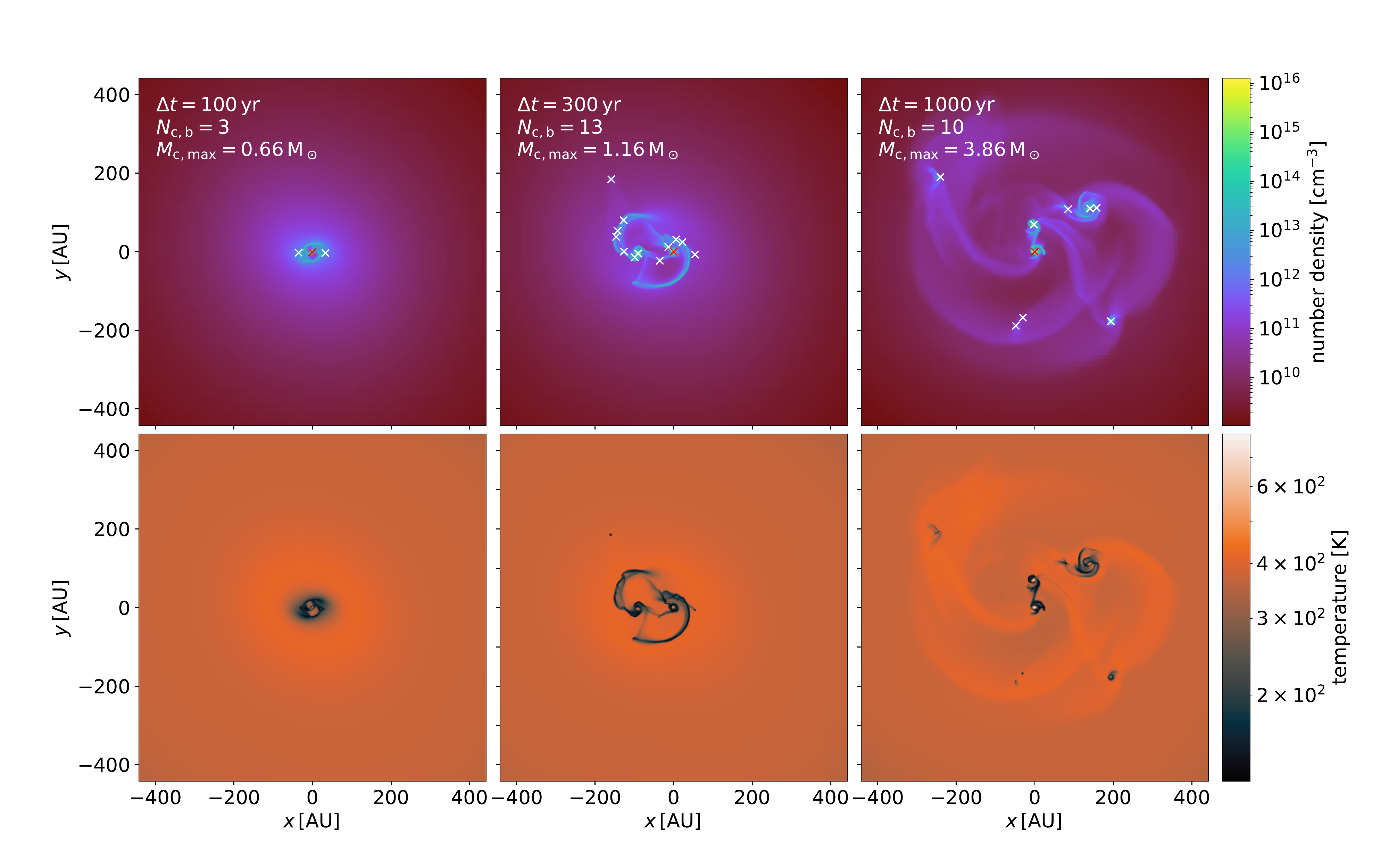}
  \caption{Same as Fig.~\ref{fig:Z0b6_projection} but for the case of $Z = 10^{-4}\,\zsun$. The clump number takes the maximum at the second snapshot $\dt = 300\,\mathrm{yr}$ (middle panels), which is later than the case of $Z = 10^{-5}\,\zsun$ (cf. Fig.~\ref{fig:Z5b6_projection}). 
  Note that the color scaling for both the density and temperature is different from Fig.~\ref{fig:Z0b6_projection}. The spatial scale presented in each panel is also slightly smaller than Figs.~\ref{fig:Z0b6_projection} and \ref{fig:Z5b6_projection}.
  }
  \label{fig:Z4b6_projection}
\end{figure*}
%-----------------------------------------------%
%------------------ Fig.6 ----------------------%
\begin{figure*}
  \includegraphics[width=1.9\columnwidth]{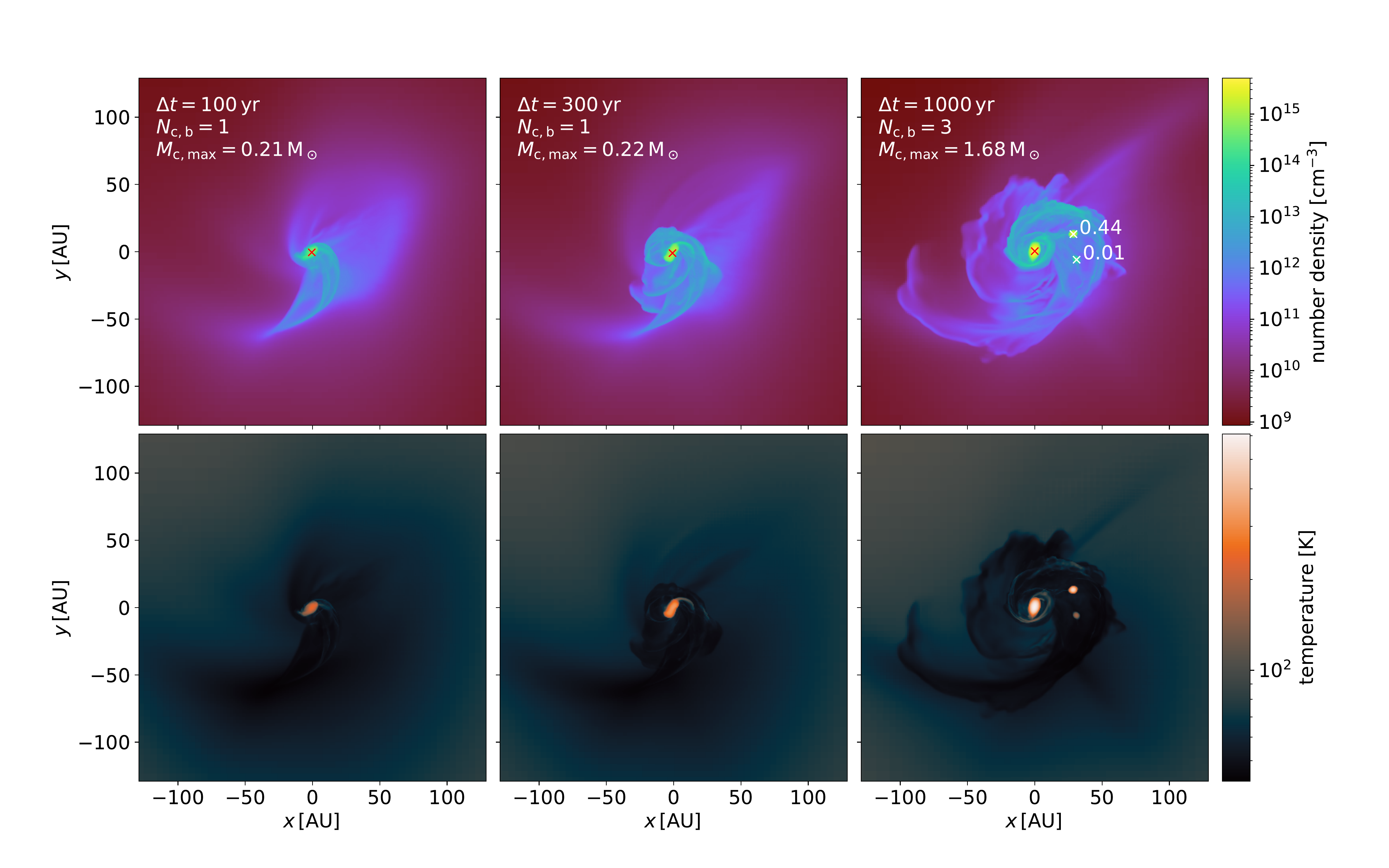}
  \caption{
    Same as Fig.~\ref{fig:Z0b6_projection} but for the case of $Z = 10^{-3}\,\zsun$.
    In the right top panel, the masses of the companion clumps are also described in the unit of $\msun$.
    Note that the color scaling for both the density and temperature is different from Fig.~\ref{fig:Z0b6_projection}. The spatial scale presented in each panel is much smaller than Figs.~\ref{fig:Z0b6_projection} and \ref{fig:Z5b6_projection}.
  }
  \label{fig:Z3b6_projection}
\end{figure*}
%-----------------------------------------------%
%------------------ Fig.7 ----------------------%
\begin{figure}
  \includegraphics[width=\columnwidth]{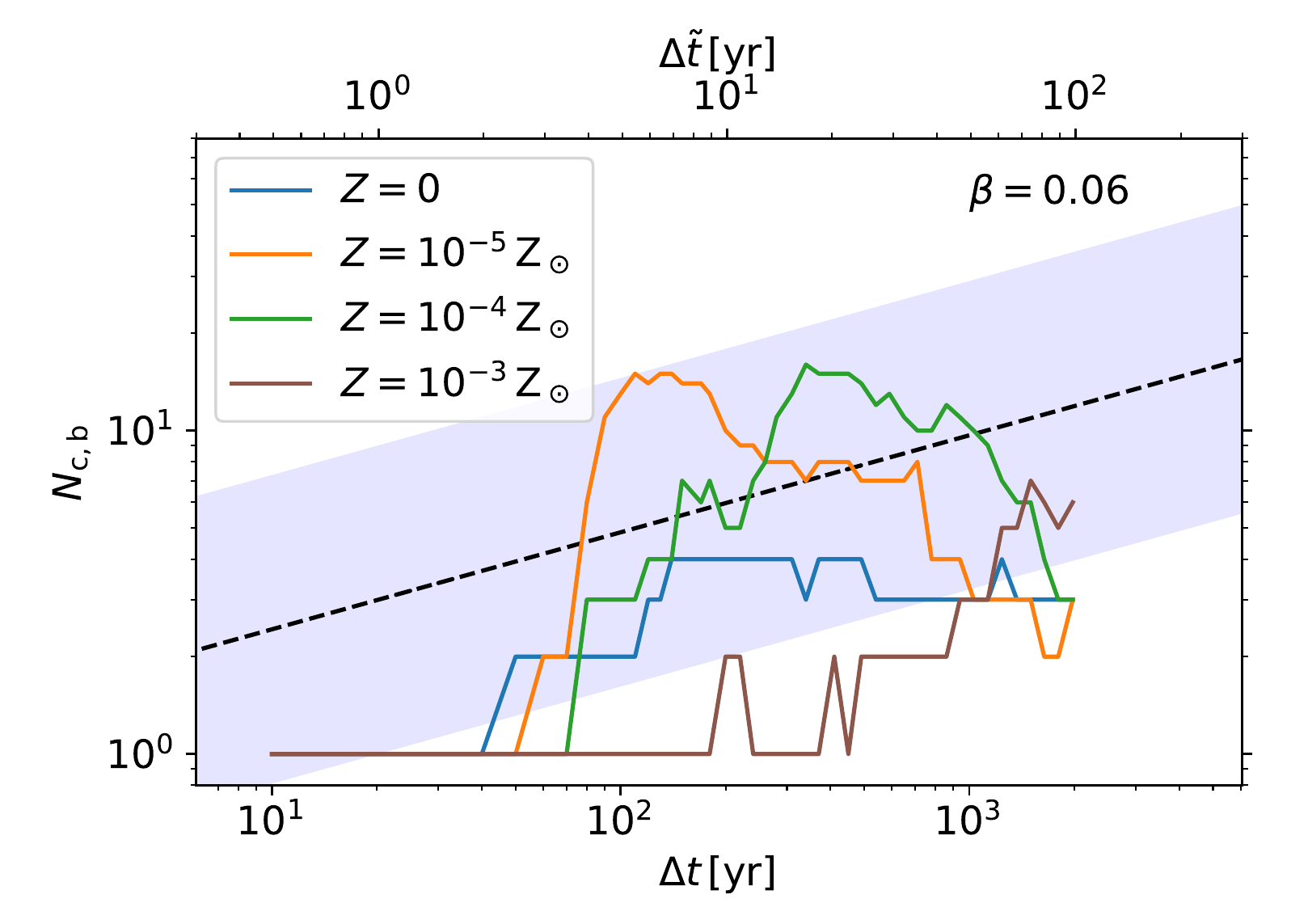}
  \caption{Time evolution of the number of gravitationally bound clumps with different metallicities. The line colors represent the cases with different metallicities, whose snapshots are presented in Figs.~\ref{fig:Z0b6_projection} - \ref{fig:Z3b6_projection}. The same cloud's rotation parameter $\beta = 0.06$ is assumed for these cases. The bottom and top horizontal axes represent the time elapsed since the first detection of a bound clump $\dt$ and $\dtt \equiv \sqrt{\nth / \nad} \dt$ (also see equation~\ref{eq:nfrag}), where $\nth = 2 \times 10^{16}\,\cc$ and $\nad = 10^{19}\,\cc$. The black dashed line represents equation \eqref{eq:nfrag2}. The blue-shaded background denotes the area within $\times 3$ and $\times 1/3$ of the fitting function, over which previous simulation results for the primordial cases are distributed as shown by \citet{Susa19}. 
}
  \label{fig:b6_susaplot}
\end{figure}
%------------------------------------------------%
%--------------------- dt=2kyr -------------------%
\begin{figure*}
\includegraphics[width=2\columnwidth]{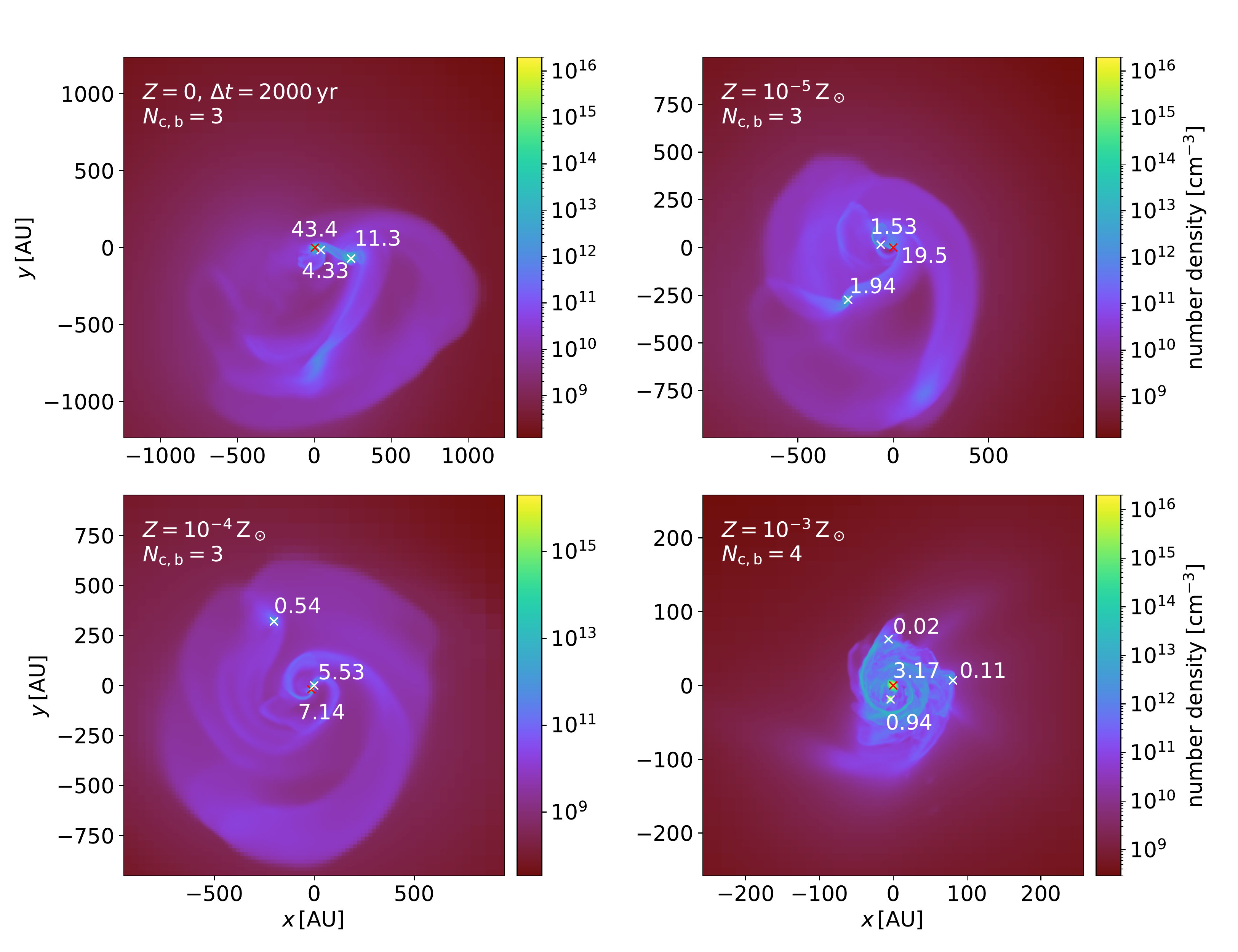}
\caption{Final snapshots at $\dt = 2 \times 10^3$~yr of the face-on projected number density maps for $\beta = 0.06$. The top-left, top-right, bottom-left, and bottom-right panels represent the cases of $Z = 0$, $10^{-5}~\zsun$, $10^{-4}~\zsun$, and $10^{-3}~\zsun$, respectively. Note that the panels cover wider areas around the most massive clump than in Fig.~\ref{fig:Z0b6_projection} and \ref{fig:Z5b6_projection} - \ref{fig:Z3b6_projection}, which show the earlier evolution for the same cases. 
The values associated with the cross symbols indicate the clump masses in the unit of $\msun$.
}
\label{fig:dt200}
\end{figure*}
%-------------------------------------------------%
%--------------------- Fig.8 ---------------------%
\begin{figure*}
  \includegraphics[width=2\columnwidth]{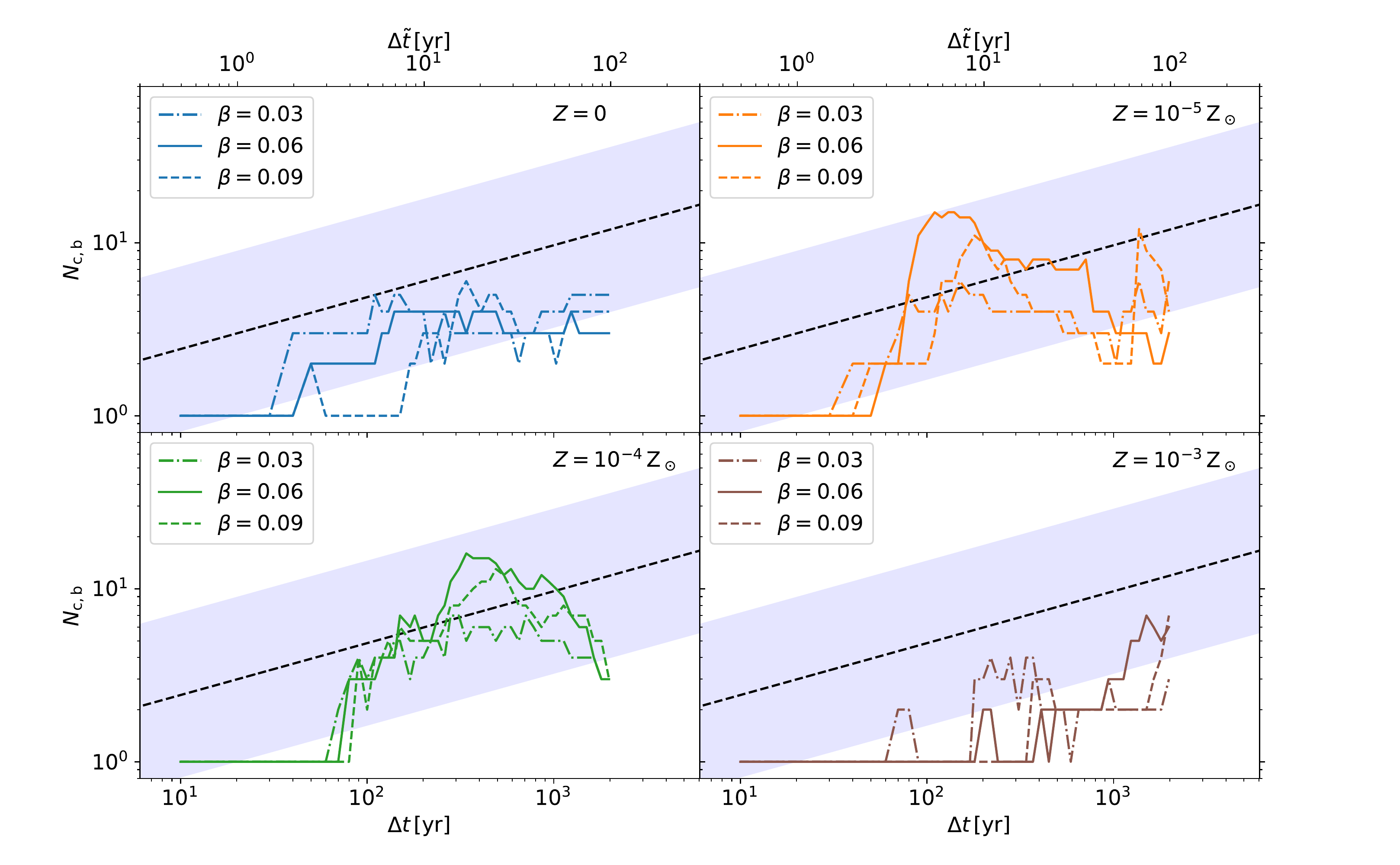}
  \caption{Same as Fig.~\ref{fig:b6_susaplot} but for varying the cloud rotation parameter $\beta$. 
  Each panel shows the evolution at a given metallicity as indicated at the upper right corner. The dot-dashed, solid, and dashed lines represent the cases with $\beta = 0.03$, 0.06, and 0.09, respectively.  
  }
  \label{fig:Z_susaplot}
\end{figure*}
%------------------------------------------------%
%-------------------- Fig.9 --------------------%
\begin{figure}
  \includegraphics[width=\columnwidth]{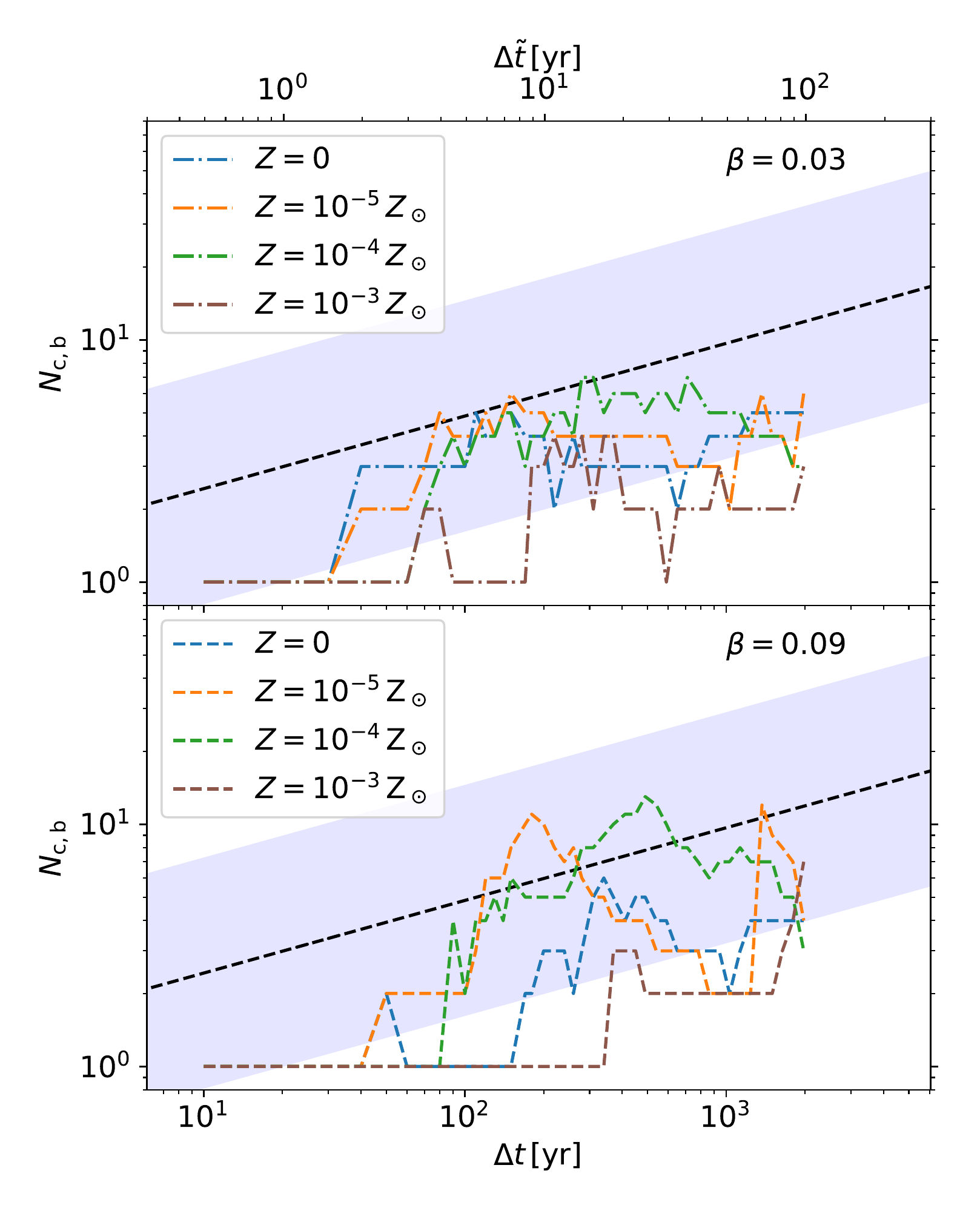}
  \caption{Same as Fig.~\ref{fig:b6_susaplot} but for the cases with $\beta = 0.03$ (upper panel) and $\beta = 0.09$ (lower panel). }
  \label{fig:b3b9_susaplot}
\end{figure}
%--------------------------------------------------%
%--------------------- Fig.10 ---------------------%
\begin{figure*}
  \includegraphics[width=2\columnwidth]{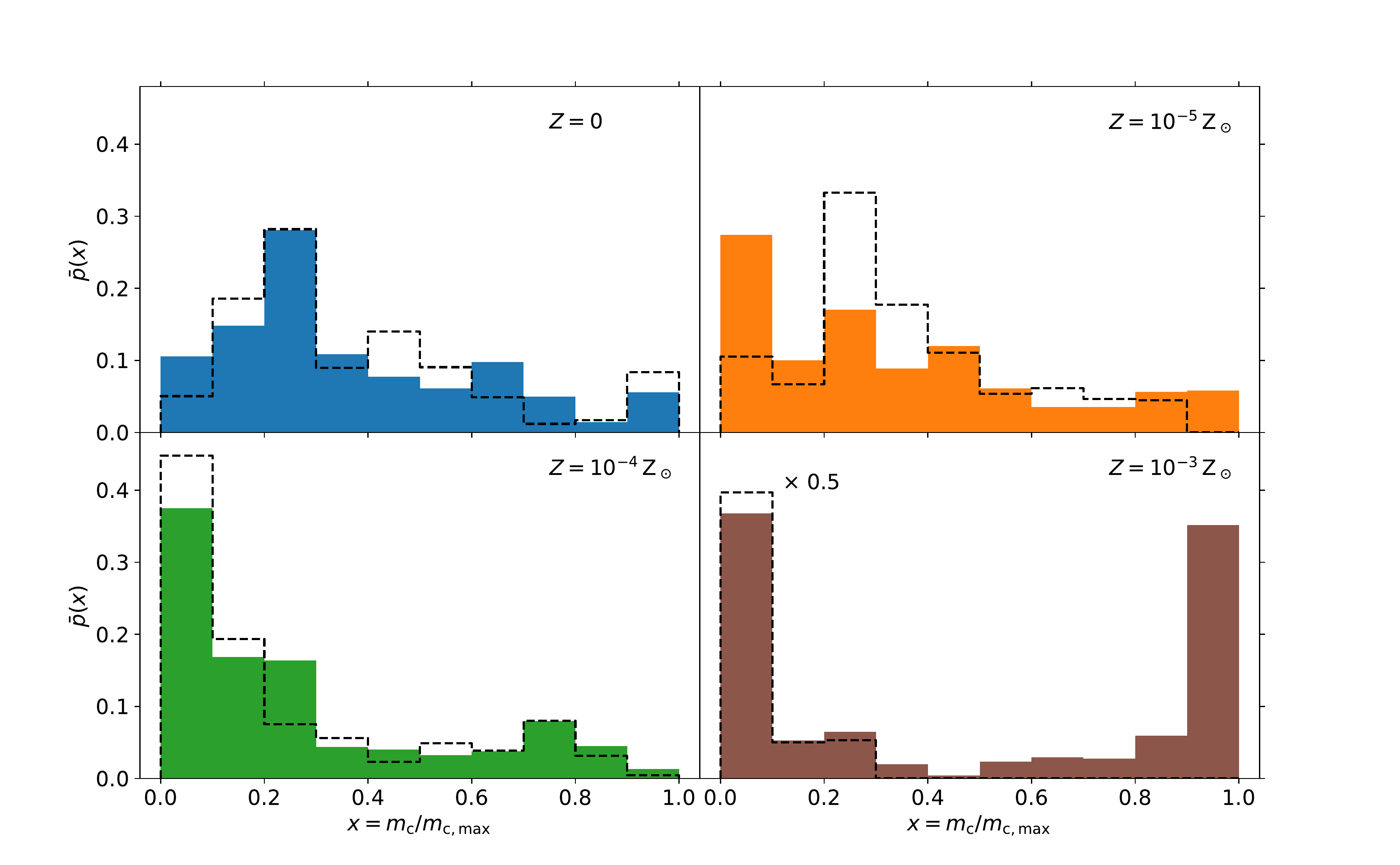}
  \caption{Normalized time-averaged mass distributions of self-gravitating companion clumps at different metallicities. The horizontal axis $x$ represents the ratio of the companion mass $\mc$ to the maximum clump mass $\mcmax$, and vertical axis represents the time-averaged probability distribution function $\bar{p}(x)$. In each panel, the filled histogram represents the probability distribution averaged over the cases with different initial cloud spin parameter $\beta = 0.03$, $0.06$, and $0.09$ at each metallicity. The black dashed line also shows the distribution only with the case of $\beta = 0.06$. In the right bottom panel for $Z = 10^{-3}\,\zsun$, plotted with the dashed line is $0.5 \times \bar{p}(x)$ so that it does not go beyond the panel
  (the apparent value at the lowest bin of $x$ is $\simeq 0.4$, but the actual value is $\simeq 0.8$.)
  }
  \label{fig:hist}
\end{figure*}
%------------------------------------------------%
%------------------Fig.11 -----------------------%
\begin{figure}
  \includegraphics[width=\columnwidth]{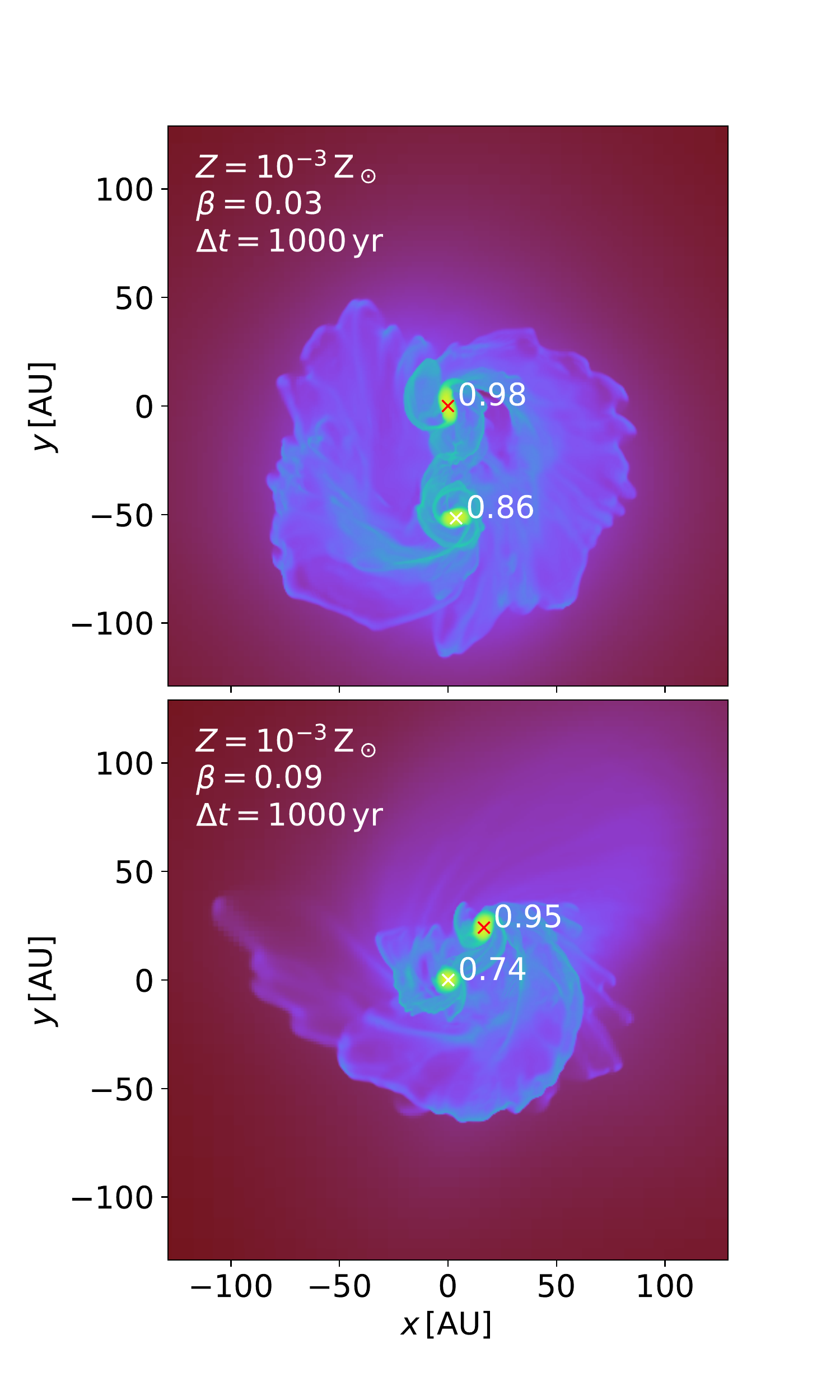}
  \caption{Comparing images of the number density maps with the different initial cloud rotation (top: $\beta = 0.03$, and bottom: $\beta = 0.09$) for $Z = 10^{-3}\,\zsun$.
  The snapshots are taken at the epoch of $\dt = 10^3\,\mathrm{yr}$ for both cases. The mass ratio of the clumps is almost unity for these cases, whereas it is small with $\beta = 0.06$ at the same epoch (cf. the top right panel in Fig.~\ref{fig:Z3b6_projection}). 
  The color scale is the same as in Fig.~\ref{fig:Z3b6_projection}.
  The mass of each clump is also described in the unit of $\msun$.
  }
  \label{fig:binary}
\end{figure}
%------------------------------------------------%
%------------------Fig.12 -----------------------%
\begin{figure*}
  \includegraphics[width=2\columnwidth]{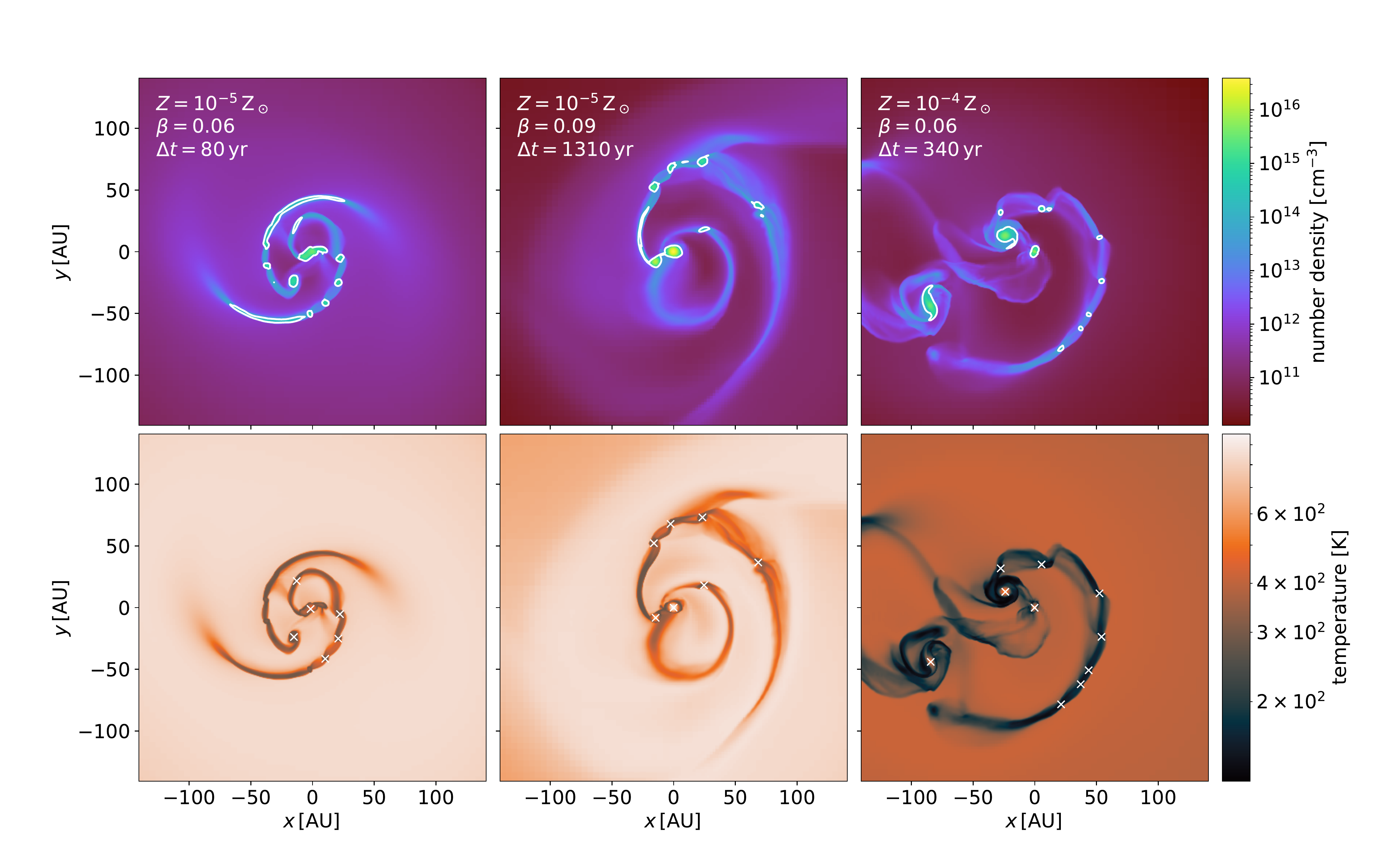}
  \caption{Featured images of the density-weighted projection maps of the number density (top) and temperature (bottom) at the epochs when the number of clumps remarkably increases due to efficient fragmentation. The left, middle, and right columns of panels represent different runs for $(Z, \beta) = (10^{-5}\,\zsun, 0.06)$, $(10^{-5}\,\zsun, 0.09)$, and $(10^{-4}\,\zsun, 0.06)$, respectively. The elapsed time $\dt$ for each snapshot is also presented in each top panel. The white lines in the density maps represent the iso-density contours at $n = 10^{14}\,\cc$. The positions of self-gravitating clumps are denoted with white crosses in the temperature maps.}
  \label{fig:fragment}
\end{figure*}
%------------------------------------------------%
%------------------ Fig.13 -----------------------%
\begin{figure}
  \includegraphics[width=\columnwidth]{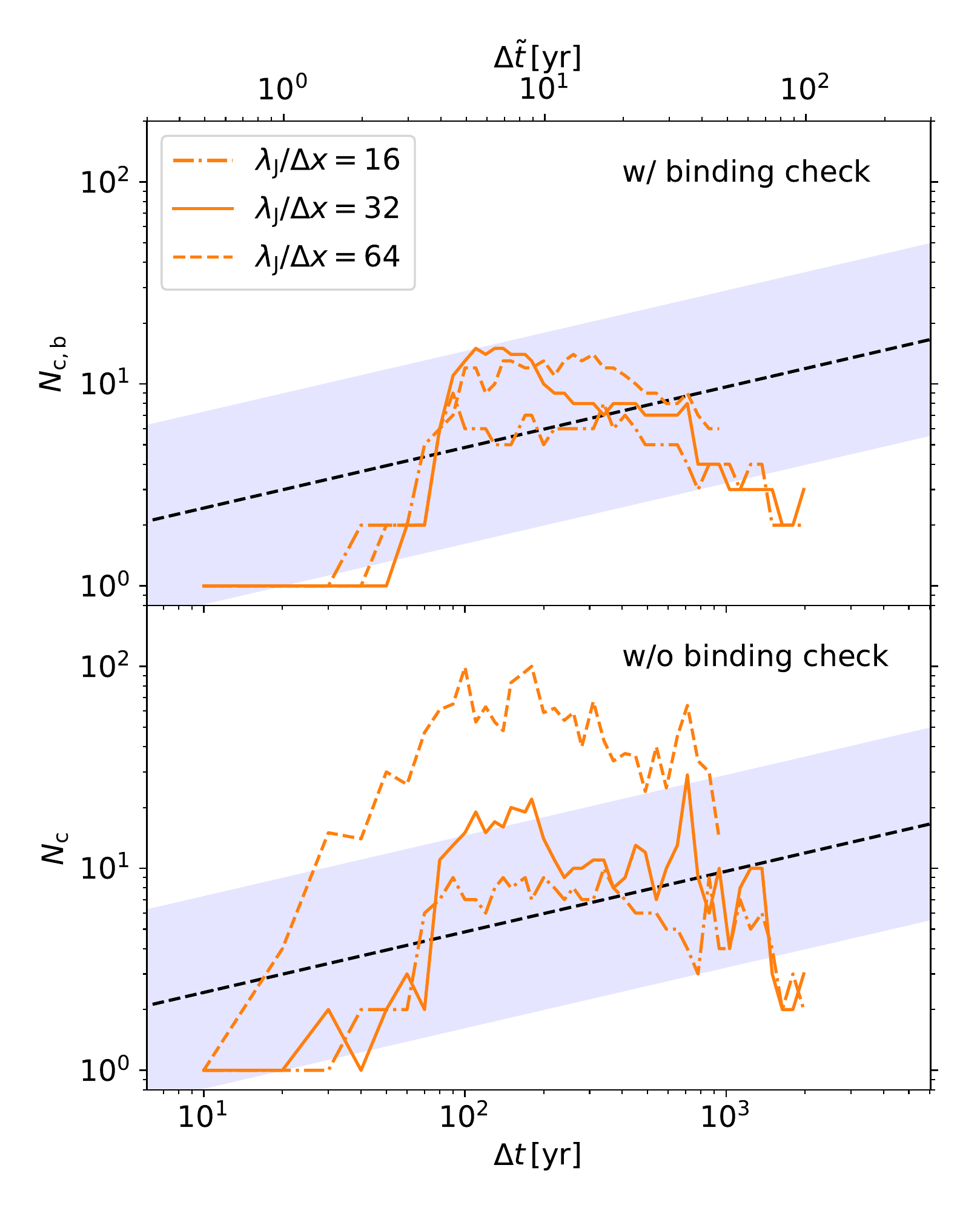}
  \caption{Time evolution of the number of clumps found by the finder with (upper) and without (lower) the gravitational binding energy check in the case of $Z = 10^{-5}\,\zsun$ and $\beta = 0.06$. In both panels, the sold line represents our standard case with the Jeans criterion $\jeansdx = 32$, i.e., the Jeans length is always resolved by at least 32 cells. The dot-dashed and dashed lines represent the less and more stringent criteria, $\jeansdx = 16$ and $64$, respectively. The green line terminates at $\dt \simeq 800\,\mathrm{yr}$ since we stop the corresponding high-resolution simulation at this point because of the heavy computational cost.
  }
  \label{fig:convergence_test}
\end{figure}
%--------------------------------------------------%

%%%%%%%%%%%%%%%%%%%
\section{Results}
\label{sec:result}
%%%%%%%%%%%%%%%%%%%

In what follows we present our simulation results. We first describe the metallicity-dependence of the evolution, considering the cases with the rotation parameter fixed at $\beta = 0.06$ in Sections~\ref{ssec:dfrag} and \ref{ssec:nfrag_evol}. We next study the effects of varying the parameter $\beta$ in Section~\ref{ssec:rot}, where the other cases with $\beta = 0.03$ and $0.09$ are presented. We finally investigate how the mass distribution of the self-gravitating clumps varies with the different metallicities in Section~\ref{ssec:massdist}. Simulation movies for all these cases are available in the (online only) supplementary data.

%-----------------------------------------------------------%
\subsection{Disc fragmentation with different metallicities}
\label{ssec:dfrag}
%-----------------------------------------------------------%

\subsubsection{Primordial case $(Z=0)$}
\label{sssec:z0_case}

Fig.~\ref{fig:Z0b6_projection} shows the face-on images of the disc fragmentation occurring for the primordial case with $\beta = 0.06$. The snapshots are taken at epochs $\dt = 100$, 300, and $10^3\,\mathrm{yr}$ after the first clump appears. The center of each panel corresponds to the most massive cell. The crosses in the density maps represent the mass centers of the clumps identified by the finder starting with $\niso = 10^{14}\,\cc$ (see Section~\ref{sec:clump_finder}). The number of gravitationally bound clumps $\Ncb$ and maximum clump mass $\mcmax$ are also presented in the upper left corner of each top panel. The earliest $\dt = 100\,\mathrm{yr}$ snapshot clearly shows the disc structure with the two spiral arms, which reflect the initial density perturbation of $m=2$ mode. We see that the accretion envelope surrounding the disc still has smooth structure in axial symmetry. The disc radially spreads as it accretes the gas coming from the envelope afterward (see also Table 3 for the quantitative assessment).
In the second $\dt = 300\,\mathrm{yr}$ snapshot, there is still the same binary found in $\dt = 100\,\mathrm{yr}$ near the center, but two more clumps also appear in the outer part of the disc after the additional fragmentation.
The latter two clumps migrate inward over a few hundred years, which is shorter than the Kepler orbital time at their birth places,
\begin{equation}
P_\mathrm{K} \simeq 1.3 \times 10^3~\mathrm{yr} 
\left( \frac{r}{200~\mathrm{au}} \right)^{3/2}
\left( \frac{M_{\rm c}}{5~\msun} \right)^{-1/2} ,
\label{eq:Kepler}
\end{equation}
where $r$ is the radial distance from the mass center. Such rapid migration has been also reported in previous studies on the primordial star formation \citep[e.g.][]{Hosokawa16}.
The migrating clumps eventually merge with the former clumps near the disc's center. 
Meanwhile, another clump appears because of the fragmentation that occurs a few $\times 100\,\mathrm{au}$ away from the center. The last $\dt = 10^3\,\mathrm{yr}$ snapshot consequently shows one clump in the outer large orbit and the other two in a central tight binary system. The binary separation is $\simeq 24\,\mathrm{au}$ in this epoch. Afterwards, there are almost always a few clumps during the evolution followed by our simulation. We also consider the evolution of the clump number in Section~\ref{ssec:nfrag_evol}, with reference to \citet{Susa19}.

%-----------------------------------------%

As shown in Fig.~\ref{fig:b6_mass}, the total mass of the clumps exceeds $50\,\msun$ after the first $2 \times 10^3\,\mathrm{yr}$, indicating the mean accretion rate of $\simeq 0.03 \msunyr$, a typical value in the primordial star formation \citep[e.g.][]{Hirano14}. The most massive clump accretes $\sim 20\,\msun$ of the gas by the epoch of $\dt = 10^3\,\mathrm{yr}$. Since the clumps represent accreting protostars, radiation emitted from such massive ones may affect the evolution of the disc. One of the key feedback mechanism is
stellar ultra-violet radiation, creating the photoionised and photodissociation regions around a protostar \citep[e.g.][]{McKee08, Hosokawa11, Hosokawa16, Fukushima20, Sugimura20}. Our simulations, where these effects are assumed to be negligible, only consider the disc fragmentation before the radiative feedback begins to operate. We note that our assumption is more justified for the low-metallicity cases shown below, where the maximum clump mass is much lower than the primordial case at the end of the simulations (see also Section~\ref{sssec:rad_fdbk}).

%-------------------------------------------------%

\subsubsection{Case of $Z = 10^{-5}\,\zsun$}
\label{sssec:z-5_case}

Fig.~\ref{fig:Z5b6_projection} presents how the disc fragmentation proceeds for the case of $Z = 10^{-5}\,\zsun$ and $\beta = 0.06$. We see that the growing disc easily fragments, forming many self-gravitating clumps. Such basic evolution may appear to be similar to the primordial case, but they are quantitatively very different. The snapshot of $\dt = 100\,\mathrm{yr}$ shows that there are more than ten fragments along the outstanding spiral structure. We have confirmed, by means of the method described in Section~\ref{sec:clump_finder}, that these objects are all gravitationally bound. The clump number of $\Ncb=13$ is much more than that for the primordial case at the same epoch, $\Ncb = 2$ (Fig.~\ref{fig:Z0b6_projection}). The bottom panel for the same snapshot shows that the spiral arm is colder than the other part of the disc, which is in stark contrast to the $Z=0$ case (see also Table 3).
The difference comes from the metallicity-dependent EOS we assume. As illustrated in Fig.~\ref{fig:n_vs_t}, the curve for $Z = 10^{-5}\,\zsun$ shows the remarkable temperature decline for $n \gtrsim 10^{12}\,\cc$ and the local minimum at $n \sim 10^{14}\,\cc$. The density and temperature along the spiral arm take these minimum values, suggesting that dust-induced fragmentation occurs. 
Interestingly, Table 3 shows that the clumps cover about 25~\% of the mass contained in the arms for both cases of $Z=0$ and $Z=10^{-5}~\zsun$ at $\dt = 300\,\mathrm{yr}$, despite the large difference in the clump numbers. 
We separately focus on this feature later in Section~\ref{ssec:fragmentation}. The subsequent $\dt = 300\,\mathrm{yr}$ and $10^3\,\mathrm{yr}$ snapshots show that the number of clumps continuously decreases with time, which differs from the primordial case. The clumps undergo the complex orbital evolution through gravitational interaction with each other, and most of them merge away during that. 
Again, we may interpret the timescale of this process as in Section~\ref{sssec:z0_case}. A clump that appears in an outer part of the disc migrates inward over the timescale comparable to or even shorter than the Kepler orbital time, $\sim 10^3$~years as indicated by equation (\ref{eq:Kepler}). Once the clump gets closer to another one, their orbital time becomes shorter, and there are more chances of the mutual interactions.
As a result, only three clumps survive at the last snapshot of $\dt = 10^3\,\mathrm{yr}$, This number is coincidentally the same as that for the primordial case at the same epoch.  

%------------------------------------------%

Figs.~\ref{fig:b6_mass} and \ref{fig:Z5b6_projection} indicate that the clumps accrete the gas of $\simeq 20\,\msun$ in total, of which the primary one dominates $\simeq 14\,\msun$ at the epoch of $\dt = 10^3\,\mathrm{yr}$. The total mass is lower than the primordial case by a factor of a few. 
%This is also due to the difference in EOS. 
The disc-star system accretes the gas from the central part of the envelope during $\dt = 2000\,\mathrm{yr}$, which is nearly equal to $\tff$ at $n \sim 10^9\,\cc$. Fig.~\ref{fig:n_vs_t} presents that the temperature at $n \sim 10^9\,\cc$ differs by a factor of two among $Z=0$ and $Z=10^{-5}\,\zsun$ cases. Given that the total accretion rate depends on the envelope temperature as $\dot{M} \propto T^{1.5}$, the difference in the mass growth histories agrees with our assumed EOS.

%-----------------------------------------------------%

\subsubsection{Case of $Z = 10^{-4}\,\zsun$}
\label{sssec:z-4_case}

Fig.~\ref{fig:Z4b6_projection} represents the case of $Z = 10^{-4}\,\zsun$ and $\beta = 0.06$. The evolution for this case quantitatively differs from both of $Z=0$ and $Z=10^{-5}\,\zsun$ cases described above. The first $\dt = 100\,\mathrm{yr}$ snapshot shows that there are three fragments along in a straight line due to the initial perturbation. 
%Unlike the $Z = 0$ case, the perturbation makes a bar structure at the center also forms a clump.
Whereas such density structure is more or less similar to the primordial case (Fig.~\ref{fig:Z0b6_projection}), the temperature distribution looks very different owing to the different EOS (Fig.~\ref{fig:n_vs_t}). 
We see a cold part at $r \lesssim 30\,\mathrm{au}$, which corresponds to the temperature decline for $n \gtrsim 10^{12}\,\cc$. The outer part shows almost the same temperature at $\simeq 400\,\mathrm{K}$, reflecting the plateau for $10^9\,\cc \lesssim n \lesssim 10^{12}\,\cc$. Subsequently, the evolution similar to $Z = 10^{-5}\,\zsun$ case continues. The snapshot of $\dt = 300\,\mathrm{yr}$ displays that there are thirteen clumps near the spiral arms. The density and temperature along the spiral arms correspond to the values at the local minimum of the EOS curve at $n \sim 10^{13.5}\,\cc$. This fact suggests that the dust-induced fragmentation yields the clumps as in $Z = 10^{-5}\,\zsun$ case, but it occurs later. The clump number slightly decreases by the last snapshot of $\dt = 10^3\,\mathrm{yr}$ because of a few merger events. They are more sparsely distributed than in the previous snapshot after complex gravitational interactions. The clump number further continues to decrease until the end of the simulation, $\dt = 2 \times 10^3\,\mathrm{yr}$ (also see Section~\ref{ssec:nfrag_evol} below). 

%---------------------------------------------------------%

We remark that in the bottom panels the clumps are more outstanding than $Z=0$ and $10^{-5}\,\zsun$ cases describe above, represented by the bright (or hot) spots surrounded by the dark (or cold) regions. We interpret this trend as follows. As seen in Fig.~\ref{fig:n_vs_t}, the EOS curves for $Z \geq 10^{-5}\,\zsun$ all converge to the same line for $n \gtrsim 10^{15}\,\cc$, where $\geff \simeq 7/5$ \citep{Omukai00}. This part corresponds to the so-called "first adiabatic core" stage \citep{Larson69}, which starts from the lower density at the higher $Z$. Since the Jeans length is in proportion to $n^{-3/10}$ with $\geff = 7/5$, the lower-density core has the larger size. 
Moreover, the disc size at a given epoch $\dt$ is systematically smaller at the higher $Z$. Recall that the disc only accretes the gas from a central part of the envelope where $\tff < \dt$. The corresponding part is more compact at the higher metallicity because the Jeans length is smaller with the lower temperature. Therefore, the higher-$Z$ disc should accrete the gas with the lower angular momentum, which explains its smaller size. The above trend becomes more prominent in the case of $Z=10^{-3}\,\zsun$ described below.

%--------------------------------------------------------%

\subsubsection{Case of $Z = 10^{-3}\,\zsun$}
\label{sssec:z-3_case}

Fig.~\ref{fig:Z3b6_projection} shows the evolution for the case of $Z = 10^{-3}\,\zsun$ and $\beta = 0.06$. Note that each panel represents the central area of $\sim 300\,\mathrm{au}$ on a side, which is much smaller than in Fig.~\ref{fig:Z4b6_projection}. In this case, the snapshots for $\dt = 100\,\mathrm{yr}$ and $300\,\mathrm{yr}$ both present only one self-gravitating clump. Although we see the spiral arms develop in the disc, it does not cause the vigorous fragmentation for $\dt \leq 300\,\mathrm{yr}$, in contrast to the lower-metallicity cases described above.  
Such a difference is well understood with Fig.~\ref{fig:n_vs_t}, where the EOS curve for $Z=10^{-3}\,\zsun$ monotonically increases for $n \gtrsim 10^{10}\,\cc$. Since the density within the disc takes $n \sim 10^{10} - 10^{13}\,\cc$, the disc temperature is nearly constant at a few $\times 10\,\mathrm{K}$. The temperature substantially increases only in the interior of an adiabatic core, i.e., for $n \gtrsim 10^{13}\,\cc$. Accordingly, the central clump is relatively very hot and large against the surrounding disc in Fig.~\ref{fig:Z3b6_projection}.  
The vigorous disc fragmentation finally starts after the epoch of $\dt = 300,\mathrm{yr}$. As a result, there are three clumps in the $\dt = 10^3\,\mathrm{yr}$ snapshot.
The bottom panel for this epoch shows that the spiral arm becomes relatively colder than the surrounding gas. Note that the density just outside the disc gradually drops as it accretes the gas from the envelope. The EOS curve for $Z = 10^{-3}\,\zsun$ has a shallow minimum at $n \sim 10^{10}\,\cc$ (Fig.~\ref{fig:n_vs_t}), whose feature appears near the outer edge of the disc in the $\dt = 10^3\,\mathrm{yr}$ snapshot.
The number of self-gravitating clumps continuously increases owing to the successive fragmentation events afterward, though followed by some merger events (see Section~\ref{ssec:nfrag_evol} and Fig.~\ref{fig:b6_susaplot} below). Note that the mergers among clumps generally occur in all the above cases, regardless of different metallicities. 

%---------------------------------------------------------------%

Fig.~\ref{fig:b6_mass} shows that the mean total accretion rate onto the clumps for $\dt = 2 \times 10^3\,\mathrm{yr}$ is as low as $\sim 10^{-3}\,\msunyr$ for $Z = 10^{-3}\,\zsun$. This is lower than that for the primordial case by a factor of $\simeq 1/0.03 \simeq 33.33$ (Section~\ref{sssec:z0_case}), which is expected with the EOS curves presented in Fig.~\ref{fig:n_vs_t}. As described in Section~\ref{ssec:setup}, the initial cloud configuration for $Z = 10^{-3}\,\zsun$ case only differs from the other cases. However, its effects on the evolution of the protostellar accretion we consider should be limited because we only focus on the central dense part with $n \gtrsim 10^9\,\cc$. 
Such dense gas appears in a late stage of the run-away cloud collapse, which converges to the same similarity solution regardless of different low-density initial states \citep[e.g.][]{Larson69, Yahil83, Omukai98}.

%----------------------------------------------------------%
\subsection{Metallicity-dependent evolution of number of self-gravitating clumps}
\label{ssec:nfrag_evol}
%----------------------------------------------------------%

Fig.~\ref{fig:b6_susaplot} summarizes the metallicity-dependent evolution of the number of clumps for the cases described in Section~\ref{ssec:dfrag}. In Fig.~\ref{fig:b6_susaplot}, we also overlay the scaling relation obtained by \citet{Susa19} for the primordial cases.\footnote{Note the different notation of our figure from Fig.~10 in \citet{Susa19}; $\dtt$ (or the upper horizontal axis) in our Fig.~\ref{fig:b6_susaplot} corresponds to their $\tau (4 \pi G \rho_{\rm ad})^{-1/2}$.
} 
Since $\nth = 2 \times 10^{16}\,\cc$ for our cases, we rewrite equation~\eqref{eq:nfrag} as
\begin{equation}
    \Ncb \simeq 4.7 \left( \frac{\dt}{100\,\mathrm{yr}} \right)^{0.3}.
    \label{eq:nfrag2}
\end{equation}
We first compare our primordial case to this relation. As seen in the figure, the corresponding blue line goes slightly below the dashed line of equation \eqref{eq:nfrag2}. Our case indicates that the clump number stays almost constant at $\Ncb \simeq 3$ for $\dt > 100\,\mathrm{yr}$, whereas equation \eqref{eq:nfrag2} predicts the monotonic increase of $\Ncb$. Our case remains within the blue-shaded area, typical scatter of simulation results previously reported by different authors, until $\dt \sim 10^3\,\mathrm{yr}$. This fact suggests that our simulation predicts relatively smaller $\Ncb$ than other studies. We note that our setup is almost the same as in \citet{Susa19}, except that the basic simulation methods are different; AMR in our study and SPH in \citet{Susa19}. In fact, no previous 3D AMR simulation results have been tested against equation \eqref{eq:nfrag}, and many of previous studies compiled in \citet{Susa19} are SPH simulations. 
A possibility is that we have missed capturing some physical processes such as turbulent fragmentation \citep[][]{Hopkins13a, Hopkins13b}. Since we aim to investigate the metallicity-dependence of the disc fragmentation, we do not further study what causes the difference in detail. We only mention that a recent study by \citet{Chon21} also shows the same trend as ours in their SPH simulation (see their Fig.~10), though their particular initial conditions may affect the evolution. In what follows, we suggest that varying the initial cloud rotation (or $\beta$) or spatial resolution does not resolve the discrepancy (see Sections~\ref{ssec:rot} and \ref{ssec:res}). 

%---------------------------------------------------%

Next, we consider the evolution in the EMP cases described in Sections~\ref{sssec:z-5_case}-\ref{sssec:z-3_case}. Fig.~\ref{fig:b6_susaplot} clearly shows that the evolution of $\Ncb$ varies with different metallicities. None of them show the evolution similar to the primordial case. Among them, however, the cases of $Z=10^{-5}\,\zsun$ and $10^{-4}\,\zsun$ show the common feature; the clump number takes the local maximum $\Ncbmax$ in an early stage, and it continues to decrease afterward. The epoch of $\Ncbmax$ is somewhat delayed with increasing the metallicity, as already mentioned in Section~\ref{sssec:z-4_case}. The evolution of $\Ncb$ for these cases qualitatively differs from the primordial case. The enhancement of $\Ncb$ is caused by the dust-induced disc fragmentation, which agrees with the semi-analytic models developed by \citet{Tanaka14}. However, our simulations suggest that fragmentation is a very sporadic process. After the vigorous fragmentation, when $\Ncb$ takes the maximum, the further fragmentation ceases for a long time, during which many clumps move around and undergo mutual gravitational interactions. Interestingly, Fig.~\ref{fig:b6_susaplot} shows that the timescale over which the clump number decreases is $\sim 10^3$ years for the both cases of $Z = 10^{-5}~\zsun$ and $10^{-4}~\zsun$. As mentioned in Section \ref{sssec:z-5_case}, this is comparable to the Kepler orbital time near the disc outer edge. The ordered spiral arm do not grow until $\Ncb$ substantially drops. As a result, $\Ncb$ for these cases becomes comparable to that for $Z=0$ for the last $10^3$ years.

%----------------------------------------------------------------%

In contrast, the case of $Z=10^{-3}\,\zsun$ displays the opposite trend; $\Ncb$ remains remarkably small until the efficient fragmentation starts at $\dt \simeq 500\,\mathrm{yr}$. The similar trend has been reported by \citet{Machida15}, who study the initial $200$ years of the protostellar accretion with various metallicities. They show that the vigorous disc fragmentation only occurs for the cases with $Z \lesssim 10^{-4}\,\zsun$, which agrees with our results if only paying attention to the early evolution. Our long-term simulation shows that $\Ncb$ starts to increase later, because of the delayed dust-induced fragmentation at $Z = 10^{-3}\,\zsun$. Although we only follow the evolution for $\dt < 2 \times 10^3\,\mathrm{yr}$, the clump number may further increase in the later stage. We also expect that the clump number then becomes variable in time because of the sporadic fragmentation events followed by mergers, as shown for the cases of $Z = 10^{-5}\,\zsun$ and $10^{-4}\,\zsun$.

%---------------------------------------------------------------------------------------%

Finally, we show the final snapshots at $\dt = 2000\,\mathrm{yr}$ for the cases examined in Fig.~\ref{fig:dt200}. 
We see that for each case the disc size has become much larger than in the earlier stages presented in Figs.~\ref{fig:Z0b6_projection} and \ref{fig:Z5b6_projection}-\ref{fig:Z3b6_projection}.
While there are a few clumps for all the cases, their total mass is larger for the lower $Z$ as indicated in Fig.~\ref{fig:b6_mass}.  We find no clear $Z$-dependencies of the relative mass distribution of the clumps, which is inevitable with looking into only one specific epoch. We return to this later in Section~\ref{ssec:massdist}, where we consider the relative mass distributions by taking the average over many snapshots every 10 years until the final epoch.

%-----------------------------------------------------------%
\subsection{Effects of varying the initial cloud rotation}
\label{ssec:rot}
%-----------------------------------------------------------%

While we have fixed the cloud rotation parameter at $\beta = 0.06$ for the models described in Sections~\ref{ssec:dfrag} and \ref{ssec:nfrag_evol}, we here consider the metallicity-dependent evolution with $\beta = 0.03$ and $0.09$. The numerical setup for these additional cases is the same as before but for varying $\beta$. Overall, the basic trend we find with $\beta = 0.06$ do not change even in such cases. Fig.~\ref{fig:Z_susaplot} summarizes the evolution of $\Ncb$ with different $\beta$ at $Z = 0$, $10^{-5}$, $10^{-4}$, and $10^{-3}\,\zsun$. Varying $\beta$ causes some scatter of the lines at each metallicity, but it looks minor compared to the effects of varying $Z$. We only notice that in the cases of $Z = 10^{-5}$ and $10^{-4}\,\zsun$ the maximum of the clump number $\Ncbmax$ is highest for $\beta = 0.06$, not for $\beta = 0.09$. We do not see a systematic trend that the more rapid initial cloud rotation leads to the more vigorous disc fragmentation. 

%----------------------------------------------------%

Fig.~\ref{fig:b3b9_susaplot} displays the metallicity-dependence of the clump number evolution with the rotation parameter fixed at $\beta = 0.03$ (upper panel) and $0.09$ (lower panel). The figure shows the similar trend to the cases with $\beta = 0.06$ (Fig.~\ref{fig:b6_susaplot}) as expected from Fig.~\ref{fig:Z_susaplot}.
We only see that the scatter of the lines with $\beta = 0.03$ is smaller than the other cases. The peak values $\Ncbmax$ at $Z = 10^{-5}$ and $10^{-4}\,\zsun$ are relatively smaller than the counterparts with the higher $\beta$. Nonetheless, the epochs of $\Ncbmax$ at a given metallicity do not shift much even if varying $\beta$. The line scatter with $\beta = 0.09$ looks more similar to that with $\beta = 0.06$ (Fig.~\ref{fig:b6_susaplot}). 

%------------------------------------------------%

We finally highlight the particular case of $Z=10^{-5}\,\zsun$ and $\beta = 0.09$. In this case, the clump number evolution is similar to the other cases of the same metallicity for $\dt \lesssim 10^3\,\mathrm{yr}$. However, the clump number abruptly rises slightly after $\dt = 10^3\,\mathrm{yr}$ and then declines only for this case. The corresponding line hence has the double peaks, at which $\Ncbmax \simeq 10$. 
Note that the horizontal axis is in logarithmic scale in Figs.~\ref{fig:Z_susaplot} and ~\ref{fig:b3b9_susaplot}, and the declining timescales after the peaks are always $\sim 10^3\,\mathrm{yr}$. What happens here is the same as that we have described in Section~\ref{sssec:z-5_case}; the dust-induced vigorous disc fragmentation followed by multiple merger events. This fact also suggests that disc fragmentation is a sporadic process, particularly for EMP cases. The evolution of the clump number should not be monotonic but very variable in time. We also feature the second event of the disc fragmentation in more detail in Section~\ref{ssec:fragmentation} later.

%-------------------------------------------------------%
\subsection{Mass distribution of self-gravitating clumps}
\label{ssec:massdist}
%-------------------------------------------------------%

We have mostly focused on the metallicity-dependence of the clump number evolution caused by the disc fragmentation, inspired by recent studies on the primordial star formation \citep[e.g.,][]{Susa19}. We here further consider variations of the clump mass distribution with different metallicities. However, recall that the total mass accreted by clumps until a given epoch significantly differs with different metallicities (Fig.~\ref{fig:b6_mass}). Since this is not caused by the metallicity-dependent nature of the disc fragmentation, we instead consider the relative mass distribution of clumps against the most massive, primary object.

%--------------------------------------------------%

 We derive the relative clump mass distribution for each run as follows. We examine all the snapshots taken every 10 years after the first appearance of a self-gravitating clump, except those where there is only a single object. For a given snapshot, we normalize each clump's mass $\mc$ by the primary's mass $\mcmax$ to build up a probability distribution function $p$ as a function of the mass ratio $x \equiv \mc/\mcmax$. The primary clump here indicates the most massive one for a given snapshot, and it is not necessarily the same throughout a simulation run.
 We divide unity into 10 equal bins and evaluate $p(x)$ per bin to draw a histogram. We derive such histograms for all the available snapshots and then take their average to derive $\bar{p}(x)$. Note that we do not include the primary clump for each $p (x)$ histogram, because otherwise a prominent peak at $x=1$ always appears in the resulting $\bar{p}(x)$. 
In this sense, obtained $p (x)$ represents the relative mass distribution of companion clumps associated with the primary one. Since we have shown that the effects of varying the rotation parameter $\beta$ is relatively minor than the metallicity-dependence (Section~\ref{ssec:rot}), we further average the cases of $\beta = 0.03$, $0.06$ and $0.09$ for each metallicity.

%-----------------------------------------------------------------%
%

Fig.~\ref{fig:hist} shows the results of our analyses. For instance, the panel for $Z=0$ shows the histogram that has the peak at $x \simeq 0.2$. This indicates that a companion clump with 20\,\% mass of the primary clump is the most typical during $\dt = 2 \times 10^3$ years, the duration of our simulations. Nonetheless, the histogram has long tails, suggesting that the companion clumps are somewhat widely distributed in mass. The panels for $Z = 10^{-5}$ and $10^{-4}\,\zsun$ also show the wide distributions with the peaks shifted to the lowest-mass bin. In these cases, the companion clumps tend to be less massive than the primary one than the primordial case. Although not very clear, the histogram for $Z = 10^{-4}\,\zsun$ is more skewed to the lower $x$ than for $Z = 10^{-5}\,\zsun$, probably reflecting the dust-induced fragmentation is delayed until the primary clump grows to become relatively massive (Section~\ref{ssec:dfrag}). The panel for $Z= 10^{-3}\,\zsun$ only shows the totally different features from the others; there are the double peaks at both ends. This is due to the variation with different $\beta$, which is remarkable only at $Z= 10^{-3}\,\zsun$. The histogram only for the case with $\beta = 0.06$ (black dashed line in the bottom right panel) shows a very high peak at the lowest bin, the trend expected from the cases of $Z = 10^{-5}$ and $10^{-4}\,\zsun$. As described in Section~\ref{sssec:z-3_case}, the single clump grows until the fragmentation starts in a late stage in this case. The other peak near $x = 1$ comes from the cases with $\beta = 0.03$ and $0.09$, where a binary system appears in an early stage at $\dt \simeq$ a few $\times 100\,\mathrm{yr}$ after the fragmentation. This early fragmentation is only weak and creates a few self-gravitating clumps at maximum. The binary then continues to grow in mass, steadily accreting the gas through a circumbinary disc. Further fragmentation does not occur for a while. The two clumps equally grow, and their mass ratio approaches to unity \citep[Fig.~\ref{fig:binary}, see also][]{Chon19}. The more vigorous fragmentation starts slightly after $\dt = 10^3\,\mathrm{yr}$ with $\beta = 0.09$ (Fig.~\ref{fig:Z_susaplot}), similar to the case with $\beta = 0.06$.

%------------------------------------------------------------%

We only observe the formation of such binary systems with twin clumps substantially more massive than the others at $Z = 10^{-3}\,\zsun$. Their typical orbital period is a few hundred years, and they complete  $~ 10$ orbits by the end of simulation runs.
In the lower-metallicity cases, more clumps temporarily appear because of the early vigorous fragmentation. However, they do not grow into the binary system as formed at $Z = 10^{-3}\,\zsun$. Many clumps undergo complex gravitational interactions that often cause mergers. A next fragmentation episode occurs when the clump number settles down to a few, as seen for the case of $Z = 10^{-5}\,\zsun$ and $\beta = 0.09$ (Section~\ref{ssec:rot}). Such a harsh environment prevents the steady mass growth of twin clumps in a binary. 
Since the efficient fragmentation begins before the end of simulations for $Z = 10^{-3}\,\zsun$, the binary system may be eventually disrupted or destroyed by gravitational interactions with other clumps. To consider the survival of the binary, we need further long-term simulations that follow the later evolution, a task for future studies.

%%%%%%%%%%%%%%%%%%%%%%%
\section{Discussion}
\label{sec:discussion}
%%%%%%%%%%%%%%%%%%%%%%%

%----------------------------------------------%
\subsection{Role of spiral arm fragmentation}
\label{ssec:fragmentation}
%----------------------------------------------%

As described in Section~\ref{sssec:z-5_case} and \ref{sssec:z-4_case}, the vigorous disc fragmentation caused by efficient dust cooling occurs particularly for the cases of $Z = 10^{-5}\,\zsun$ and $10^{-4}\,\zsun$. We here further investigate how this process develops in more detail. 
The disc fragmentation has been intensely studied, mostly in the context of the present-day star and planet formation \citep[see][for a review, and references therein]{Kratter16}. 
Previous studies have already provided a key concept that the gravitational instability of the spiral arms essentially represents the disc fragmentation \citep[e.g.][]{Takahashi16,Brucy21}.
\citet{Takahashi16} show that the linear stability analysis of a rotating ring or filament well describes the spiral-arm instability, with thorough comparisons to their 2D simulations. Whereas \citet{Takahashi16} specifically consider the fragmentation of a non-accreting massive protoplanetary disc, \citet{Inoue18} show that the similar concept is applicable for the galaxy-formation simulations in 3D. \citet{Inoue20} further apply the analyses to the cosmological simulation of the primordial star formation performed by \citet{Greif12}. They show that the fragmentation of a rapidly accreting circumstellar disc demonstrated by \citet{Greif12} is essentially the same process.

%-----------------------------------------------------------------%

Figs.~\ref{fig:Z5b6_projection} and \ref{fig:Z4b6_projection} have already suggested that the temperature and density along the spiral arms roughly correspond to the values at the local minima of the EOS curves at $n \sim 10^{14}\,\cc$ (Fig.~\ref{fig:n_vs_t}), where $\geff = 1$.
Fig.~\ref{fig:fragment} now further displays featured images just after the disc fragmentation for the same cases as in Figs.~\ref{fig:Z5b6_projection} and \ref{fig:Z4b6_projection} in the left and right columns. These snapshots are taken at $\dt = 80\,\mathrm{yr}$ and $340\,\mathrm{yr}$ for the former and latter cases, which correspond to the epochs of $\Ncbmax$ in Fig.~\ref{fig:b6_susaplot}.  
The middle column shows the snapshots at the much later stage of $\dt = 1310\,\mathrm{yr}$ for the case of $Z = 10^{-5}\,\zsun$ and $\beta = 0.09$. 
This corresponds to the second peak of $\Ncb$, the second episode of the disc fragmentation that happens after the clump number decreases down to a few (Fig.~\ref{fig:Z_susaplot}). 

%-------------------------------------------------------------------%

We find the common feature for all the cases presented in Fig.~\ref{fig:fragment}. The geometrically thin spiral arms develop, and it is stretched to become the filamentary structure. At first glance, we recognize that the self-gravitating clumps are distributed well along the spiral arms and that the iso-density contours delineate them at $n = 10^{14}\,\cc$. 
The dispersion relation derived by the linear stability analysis predicts that the most unstable wavelength is $\simeq 4$ times larger than the arm width \citep{Takahashi16}. The distribution of the clumps shown in Fig.~\ref{fig:fragment} apparently agrees with the picture of the spiral-arm instability. We can further interpret why such vigorous fragmentation is a sporadic process. A fragmentation event produces $\sim 10$ clumps at once, and the violent motion of these clumps prevents the growth of the spiral arms that cover the whole disc. It is after the clump number settles down to a few that the large-scale spiral arms start to grow, leading to the next fragmentation episode.

%-----------------------------------------------------------------------------------%
\subsection{Low-metallicity dust-induced fragmentation: further challenges}
\label{ssec:uncertain_dust_frag}
%-----------------------------------------------------------------------------------%

We have demonstrated that the dust cooling induces the efficient disc fragmentation for the EMP cases. This is an aspect of the dust-induced fragmentation, a paradigm that describes the transition between the first and second generation stars in the early universe. 
Peculiar composition of the Galactic EMS star SDSS J102915+172927 \citep{Caffau11} has been proposed as a possible signature of the dust-induced fragmentation \citep[][]{Klessen12, Schneider12b,Chiaki14,Bovino16}. However, there are still challenges to be investigated in further studies. 
For instance, we use the pre-calculated barotropic EOS taken from \citet[][see also Fig.~\ref{fig:n_vs_t}]{Omukai05}, which suffers from limitations. Relying on the tabulated EOS is a rough approximation as already mentioned in Section~\ref{ssec:EOS}. 
The better treatment is fully solving the energy equation and separately determining the gas and dust temperatures \citep[e.g.][]{Dopcke11,Dopcke13}. Since such an improvement increases the computational cost, it is often trade-off with extending the duration of the evolution one follows.

%-------------------------------------------------------------------------------------------%

Moreover, \citet{Omukai05} assume the same dust size distribution and composition as in the solar neighborhood \citep{Mathis77}, which may be inapplicable to the EMP cases. Dust properties in the early universe should differ from the local universe, as metal-free Type-II and pair-instability supernovae (SNe) explosions presumably dominate the dust production. Theoretical studies predict that such "first dust" grains have relatively small sizes \citep[][]{Todini01, Nozawa03, Schneider04}.
3D SPH simulations by \citet{Tsuribe06} demonstrate that the dust-induced fragmentation occurs during a EMP cloud collapse assuming the dust produced by metal-free pair-instability SNe. 
The metal depletion factor, or the mass fraction of metals depleted onto dust grains, depends on the efficiency of the destruction process or how successfully grains survive from reverse shocks in SN remnants \citep[][]{Bianchi07, Nozawa07}. \citet{Schneider06, Schneider12} update the one-zone models by \citet{Omukai05}, incorporating such theoretical predictions for the first dust grains. They suggest that the dust-induced fragmentation should generally occur while varying the dust properties shifts the local minima of $\rho$-$T$ curves. 
\citet{Schneider12} propose that the minimum dust-to-gas mass ratio for the fragmentation is more essential than the critical metallicity, considering the uncertainties in the metal depletion factor \citep[see also][]{Bovino16}. 
We also note that possible dynamical segregation between the dust and gas may cause large fluctuations of the dust-to-gas mass ratio within a galaxy, even if the metallicity is almost homogeneous \citep{Hopkins17, Fukushima18}. Studying the effects of these variable dust properties on disc fragmentation is still to be done. Considering the dependencies on dust-to-gas mass ratio, rather than the metallicity, is also suitable for that purpose. 

%---------------------------------------------------------------------%

Grain growth processes in the dense interstellar medium, such as accretion and coagulation, have drawn attention to explain the rapid dust enrichment in galaxies \citep[e.g.][]{Asano13}. 
Whereas the efficiency of these processes are still in debate \citep{Ferrara16}, recent studies suggest that these additional processes may explain the observed dust content in high-redshift and local galaxies \citep[][]{Mancini15,Schneider16,Zhukovska16,Aoyama17,Ginolfi18}. Although the grain growth is ineffective for the EMP cases on the galactic scale, it operates to modify the EOS curve at $n \gtrsim 10^{10}~\cc$ during the collapse of an individual star-forming cloud \citep{Nozawa12,Chiaki13,Chiaki15}. 3D simulations by \citet{Chiaki16} and \citet{Chiaki20} have demonstrated that the grain growth operates to enhance the dust-induced fragmentation.\footnote{\citet{Chiaki19} show that the the grain growth is ineffective when assuming $13~\msun$ SN progenitor star, while \citet{Chiaki16} and \citet{Chiaki20} assume a $30~\msun$ progenitor star.} However, they only follow the evolution for the initial $< 100$ years of the protostellar accretion stage. If the grain growth efficiently operates later, it may further promote the disc fragmentation in the EMP cases.

%-------------------------------------------------%
\subsection{Effects of varying Jeans criterion}
\label{ssec:res}
%-------------------------------------------------%

We have imposed the condition that the Jeans length $\jeans$ must be resolved by at least 32 cells, i.e., $\jeansdx = 32$, for the simulations presented above. As described in Section~\ref{ssec:setup},  we also perform additional simulations with varying the "Jeans criterion" as $\jeansdx = 16$ and $\jeansdx = 64$. We consider the case of $Z = 10^{-5}\,\zsun$ and $\beta = 0.06$, where the vigorous fragmentation occurs in an early stage (Section~\ref{sssec:z-5_case}), for such experimental runs. 

%------------------------------------------------------------%

Fig.~\ref{fig:convergence_test} shows the time evolution of the number of clumps with different criteria. The upper panel shows the evolution of the self-gravitating clump number $\Ncb$. We see that the basic evolution does not change even if varying the Jeans criteria, though the peak value $\Ncbmax$ is slightly reduced with $\jeansdx = 16$. 
There is a common trend that $\Ncb$ gradually decreases after taking peak values at $\dt \sim 100\,\mathrm{yr}$ because of many merger events. 
The number of surviving clumps eventually converges to almost the same value at $\dt \sim 10^3\,\mathrm{yr}$. The figure shows that the evolution is particularly similar if $\jeansdx > 32$, and it well demonstrates the numerical convergence of our results.   

%-----------------------------------------------------------------%

We also consider the role of checking whether a clump is gravitationally bound or not (Section~\ref{sec:clump_finder}). The lower panel shows the evolution of the clump number without the binding check, $\Nc$. We see the larger variations of the lines than in the upper panel. We detect almost always more clump candidates with the more stringent criterion. This is not surprising because the smaller transient structure of the disc is resolved with the higher-resolution simulation realized with the stringent Jeans criterion. In particular, the number of such transient structure is more numerous by an order of magnitude than the self-gravitating clumps with $\jeansdx = 64$. This suggests how critical the binding check is for counting the clump number. One may significantly overestimate the number if misidentifying the transient structure. 

%------------------------------------------------------------------%

We note that the criterion of minimum spatial resolution generally changes with situations. If we consider an initially turbulent star-forming cloud, it is critical how well one resolves the turbulent eddies for subsequent evolution. \cite{Kritsuk07} and \cite{Federrath11} show that changing the Jeans criterion directly impacts the simulation results for such cases. \cite{Meece14} demonstrate that at least 64 cells per Jeans length are necessary to capture the turbulence using the {\it Enzo} code.  Moreover, following the thermal and chemical evolution of a shocked cooling layer also needs a very high resolution. Indeed, \citet{Turk12} and \citet{Sharda21} show that for the primordial case changing the resolution affects the thickness of a compressed layer bounded by accretion shocks created around a circumstellar disc.

%------------------------------------------------------%
\subsection{Additional effects to be considered}
\label{ssec:caveat}
%------------------------------------------------------%

To isolate the possible metallicity dependence of the disk fragmentation, we have neglected several physical processes that may play important roles. We here discuss such additional effects to be considered in further studies.

\subsubsection{Turbulence}

Observations suggest that some levels of turbulence are everywhere in nearby star-forming regions \citep{Elmegreen04, Heyer15}, and it is believed to play a pivotal role in the present-day star formation \citep[][]{Scalo04, MacLow04, Hennebelle12}. Turbulent motion easily creates density perturbations and causes the fragmentation of a cloud \citep[e.g.][]{Girichidis20}, and it regulates the angular momentum of the gas falling onto a protostellar disc \citep[e.g.][]{Zhao20}. Cosmological simulations suggest turbulence should also be present in EMP star-forming sites in the early universe, owing to dynamical metal enrichment processes caused by SN explosions \citep[e.g.][]{Ritter12, Smith15, Chiaki18}.

%---------------------------------------------------------------------%

Whereas a gravitationally unstable disc generally involves turbulence, turbulent motion present in an earlier stage, even before the onset of the cloud collapse, also affects the disk fragmentation. 
Some previous simulations show that for the primordial case different realizations of the initial turbulence result in the stochastic nature of the disk fragmentation \citep[][]{Riaz18, Wollenberg20, Sharda20}. 
These suggest that a sufficiently large number of simulation runs are necessary to examine the effect of the initial turbulence on the disk fragmentation. 
Such systematic studies are still limited for low-metallicity cases.

%-------------------------------------------------------------------%

It is still uncertain how strong initial turbulence smears out the metallicity dependencies in the fragmentation process. \citet{Meece14} study the effects of systematically varying the metallicity and the initial degree of turbulence on the fragmentation during the cloud collapse before the maximum density exceeds $10^{10}~\cc$. 
They show that there is still a tendency for dust cooling to promote fragmentation above the critical metallicity, although it becomes less pronounced with stronger initial turbulence. \citet{Chon21} perform the much longer-term simulations starting with low-metallicity turbulent clouds to study the fragmentation that occurs after the formation of protostars. Although not varying the strength of the initial turbulence, they confirm that the fragmentation is more efficient at higher metallicity.

\subsubsection{Magnetic Fields}

%--------------------------------------------------------%

The presence of the magnetic fields should also affect the disk fragmentation, depending on their strengths. 
Previous studies on the primordial star formation have intensively investigated the amplification of magnetic fields owing to the turbulent dynamo, by means of analytic consideration \citep[e.g.][]{Tan+Blackman04,Schober12,Xu16,Latif16,Mckee20} and numerical simulations \citep[e.g.][]{Schleicher10,Sur10,Federrath11,Turk12,Sharda21}. 
Therefore, one should consider both the turbulence and magnetic fields simultaneously, particularly on the primordial star formation, where only weak seed fields may be available before the onset of the cloud collapse. 
\cite{Sharda20} perform a large set of magnetohydrodynamic (MHD) simulations to study the interplay between the turbulence and magnetic fields in the disc fragmentation. They show that the magnetic fields operate to suppress disc fragmentation. However, the turbulence makes the evolution very chaotic, resulting in a large scatter in the system's clustering properties, such as the stellar multiplicity.
Several authors also show that magnetically driven outflows appear if the fields are strong enough \citep[][]{Machida06,Sadanari21}. The outflow regulates mass and angular momentum supplies onto a disc \citep{Matzner00,Machida13b}, which accordingly affects the disc fragmentation.

%---------------------------------------------------------%
% low-metallicity cases, non-ideal MHD effect 

Extending the above consideration on the primordial cases to the low-metallicity star formation is underway. MHD simulations by \citet{Peters14} also demonstrate that the presence of the magnetic fields more or less suppress the disc fragmentation also for $0 < Z \leq 10^{-4}~\zsun$. 
We finally note that the above studies have assumed the ideal MHD approximation, which is not necessarily valid for the low-metallicity cases \citep[][]{Susa15,Nakauchi19, Nakauchi21}. 
Non-ideal MHD simulations that consistently solve the dissipation of the magnetic fields should reveal the realistic evolution involving both the disc fragmentation and outflow launching \citep[][]{Higuchi19}.

\subsubsection{Radiative Feedback}
\label{sssec:rad_fdbk}

We have neglected the protostellar radiative feedback in our simulations. We expect our treatment to be valid for the EMP cases, where the maximum mass of clumps (or protostars) is $\lesssim 20~\msun$ throughout the simulations.  However, we here assume that, for a star with a given large mass, the radiative feedback in the EMP cases is not much more powerful than in the primordial case. 

%---------------------------------------------------------------%

Consider the possible regulation of the mass supply onto a star-disc system from a surrounding envelope by the protostellar feedback. Given that the accreting gas contains some amount of dust grains, the radiation pressure exerted on the grains pushes away the gas via dynamical coupling \citep["radiation-force" feedback, e.g.][]{Kahn74, Wolfire87}. This qualitatively differs from the UV feedback postulated for the primordial case, which is driven by gas pressure enhanced by the photoionization \citep[e.g.][]{Omukai02, McKee08}. Therefore, these different types of radiative feedback jointly work except for the primordial case, where there are no dust grains. 

%----------------------------------------------------------------%

\cite{Hosokawa09} analytically estimate that the radiation-force feedback becomes effective only for $Z \gtrsim 10^{-2}~\zsun$, and that the photoionization feedback is the primary process for the EMP cases. 
Whereas \citet{Hosokawa09} assume the spherical accretion, \citet{Tanaka18} consider the disc accretion and obtain the qualitatively similar results using the semi-analytic modeling. 
Radiation-hydrodynamic (RHD) simulations assuming the 2D axial symmetry also confirm that the interplay between the radiation-force and photoionization feedback only appears for $Z \gtrsim 10^{-2}~\zsun$ \citep[][]{Kuiper18,Fukushima20}.
Future 3D RHD simulations are awaited to reveal the more realistic evolution involving the disc fragmentation, and they are also necessary to consider possible connections between the present-day, EMP, and primordial high-mass star formation.  The disc fragmentation under the interplay between the radiative feedback, magnetic effects, and turbulence has been studied only for the present-day case \citep[e.g.][]{Rosen20}.

%----------------------------------------------------------------------------%
\subsection{Possible connections to the further high-metallicity cases}
\label{ssec:highz}
%----------------------------------------------------------------------------%

Whereas we have only considered the EMP cases with $Z \leq 10^{-3}\,\zsun$, other authors have studied the disk fragmentation at the further high metallicities, most intensively at $Z = \zsun$ \citep[see][for a recent review]{Zhao20}. 
Recent studies on the present-day star formation have already investigated the effects discussed in Section~\ref{ssec:caveat}, or the disk fragmentation with the initial cloud-scale turbulence \citep[e.g.][]{Goodwin04,Offner10,Walch12,Tsukamoto13b}, and with magnetic fields \citep[e.g.][]{Hennebelle08,Inutsuka10,Machida11,Tsukamoto15a,Tsukamoto15b,Hennebelle16, Wurster18}. The efficiency of the magnetic braking varies with the angle between the magnetic field lines and the cloud's rotation axis \citep[][]{Matsumoto04,Price07,Joos12,Hirano20}, and the random turbulent fields easily cause the misalignment \citep{Joos13}. Therefore, the effects of the turbulence and magnetic fields might be inseparable \citep[see also][]{Seifried12}.

%------------------------------------------------------------------------------------%

Since the effects described above remain to be studied for the low-metallicity star formation, we compare our results to the earlier simulations assuming the idealized initial setup of the rigidly rotating unmagnetized clouds \citep[][]{Walch09,Tsukamoto11}. They show that the disc fragmentation only rarely occurs at $Z = \zsun$ while a massive and gravitationally unstable disc often appears.
It contrasts our EMP cases where the vigorous disc fragmentation provides $\sim 10$ self-gravitating clumps. 
Although less explored, the knowledge on the disc fragmentation at $Z \sim 0.1-0.01~\zsun$ is indispensable to bridge the gap from our EMP cases. \cite{Machida15} systematically study the disc fragmentation for a full range of the metallicities $0 \leq Z \leq \zsun$, demonstrating that the evolution looks similar for $Z \gtrsim 10^{-3}~\zsun$; the disc fragmentation hardly occurs during the initial $\sim 100$ years. \cite{Vorobyov20} follow the long-term ($> 10^5$ years) evolution for $Z \geq 10^{-2}~\zsun$ cases with 2D simulations under the thin disc approximation. They find that the disc generally fragments later and that reducing the metallicity promotes the fragmentation. \citet{Meru10} predict such $Z$-dependence because the low-metallicity disc cools efficiently with the lowered opacity.   
\citet{Bate19} shows the similar trend in larger-scale simulations of the star cluster formation, reporting that the more efficient cooling promotes the formation of close binaries at the lower metallicities.

%----------------------------------------------------------------------%

The strong protostellar radiative feedback such as discussed in Section \ref{sssec:rad_fdbk} is ineffective in the low-mass ($\sim 1~\msun$) star formation, typical at $Z \sim \zsun$. However, many simulations have demonstrated that the irradiation heating by low-mass accreting protostars operates to suppress the disc fragmentation \citep[][]{Offner09, Meru10, Bate12, Bate18}. \cite{Smith11} show that for the primordial star formation this effect is inefficient and only slightly delays the disc fragmentation. These findings agree with \citet{Omukai10}, who evaluate that the irradiation heating is only effective in preventing the fragmentation for $Z \gtrsim 10^{-3}~\zsun$ \citep[see also][]{Sharda21b}. The protostellar irradiative heating, if considered, may further reduce the number of clumps in our $Z = 10^{-3}~\zsun$ cases, which is the least among the cases examined.

%%%%%%%%%%%%%%%%%%%%%%%
\section{Conclusions}
\label{sec:conclusion}
%%%%%%%%%%%%%%%%%%%%%%%

We have studied the gravitational fragmentation of accreting circumstellar discs with various metallicities, by performing a suite of 3D hydrodynamic simulations using the adaptive mesh refinement code {\it Enzo}. We model the metallicity-dependent EOS of the gas using pre-calculated barotropic EOS at $Z=0$, $10^{-5}$, $10^{-4}$, and $10^{-3}\,\zsun$. A simulation run begins with an idealized rotating cloud characterized by the spin parameter $\beta$. 
We assume that there are no turbulence and magnetic fields in the initial state for simplicity. We have followed the evolution from the early collapse to the subsequent accretion stage. In particular, we have investigated the long-term evolution in the late accretion stage for $\dt = 2 \times 10^3\,\mathrm{yr}$, which is longer by an order of magnitude than in previous relevant studies \citep[][]{Machida15,Chiaki20}. Following these studies, we do not use the sink particle method but stiff EOS to represent accreting protostars. We ignore the protostellar radiative feedback because the duration we follow is still before it substantially affects the evolution. 
We have further studied the effects of varying the cloud rotation parameter $\beta$ and the so-called Jeans criteria performing the additional simulations. 

%---------------------------------------------------------%

Our simulations show that the disc fragmentation occurs for all the examined cases, regardless of the metallicity. However, the resulting evolution shows the clear metallicity dependence.
We have paid special attention to the evolution of the number of self-gravitating clumps formed by the fragmentation, $\Ncb$. The primordial case shows that the fragmentation, often followed by clump mergers, steadily continues until the end of the simulation. The clump number stays almost constant at a few during that, while it does not monotonically increase as suggested by \citet{Susa19}.
In contrast, the evolution of $\Ncb$ becomes variable in time with a tiny amount of metals. The vigorous fragmentation caused by efficient dust cooling occurs in an early stage at $Z = 10^{-5}$ and $10^{-4}\,\zsun$, as predicted by \citet{Tanaka14} using 1D semi-analytic modeling. The clump number temporarily rises to $\Ncb \sim 10$, but it continuously decreases as many clumps merge within $2000$~yr. The vigorous fragmentation tends to occur later with the higher $Z$, reflecting that the dust-induced fragmentation is most efficient at the lower density. At $Z = 10^{-3}\,\zsun$, the clump number is smallest until the efficient fragmentation eventually starts at $\dt \sim 10^3\,\mathrm{yr}$.
In all the cases, the clump number settles down to a few by the specific epoch of $\dt = 2 \times 10^3$ years after such very different evolution. The above picture does not change even if varying the cloud's initial rotation parameter $\beta$. 

%-----------------------------------------------------------%

We have also analyzed the simulation data to investigate the time-averaged relative mass distribution of the clumps for all the examined cases. The mass distribution also shows a systematic trend; the companion clumps become relatively less massive than the primary or most massive one with increasing $Z$. This reflects the metallicity-dependent evolution described above, i.e., the primary clump has more time to accrete the gas until the vigorous fragmentation starts at the higher $Z$. On top of this trend, there is another striking difference in the mass distribution at $Z = 10^{-3}\,\zsun$; the other peak near the high-mass end, representing a binary system with twin clumps substantially more massive than the others. We have shown that such a characteristic system grows through steady accretion from a circumbinary disc, during which the system is not disturbed by other clumps. It seems that this hardly occurs at $Z = 10^{-5}$ and $10^{-4}\,\zsun$ because many clumps produced by the early dust-induced fragmentation continue to interact with each other violently. Although our current simulations only follow the initial $2 \times 10^3$ years of the protostellar accretion stage, the result suggests how massive and equal-mass binaries form in low-metallicity environments.

\section*{Acknowledgements}

We thank Sunmyon Chon, Kazuyuki Sugimura, Ryoki Matsukoba, Gen Chiaki, Shigeki Inoue, Naoki Yoshida, Kazuyuki Omukai, and Hajime Susa for useful discussion and comments. 
We also thank an anonymous reviewer for his/her constructive comments. This work is financially supported by the Grants-in-Aid for Basic Research by the Ministry of Education, Science and Culture of Japan (17H06360, 19H01934: T.H.). The numerical simulations were performed on the Cray XC50 (Aterui II) at the Center for Computational Astrophysics (CfCA) of National Astronomical Observatory of Japan. The simulation results are analyzed using the visualization toolkit for astrophysical data YT \citep{yt}.

%%%%%%%%%%%%%%%%%%%%%%%%%%%%%%%%%%%%%%%%%%%%%%%%%%
\section*{Data Availability}
The data underlying this article will be shared on reasonable request to the corresponding author. Movies of the simulations can be found at the following link: \url{https://www.youtube.com/playlist?list=PLy0BOLTBcHhaYoWvkp5zZ982amm4QcRaK}



\bibliographystyle{mnras}
\bibliography{biblio} % if your bibtex file is called example.bib

%\begin{thebibliography}{99}
%\end{thebibliography}

%%%%%%%%%%%%%%%%%%%%%%%%%%%%%%%%%%%%%%%%%%%%%%%%%%

%%%%%%%%%%%%%%%%% APPENDICES %%%%%%%%%%%%%%%%%%%%%

%\appendix

%If you want to present additional material which would interrupt the flow of the main paper, it can be placed in an Appendix which appears after the list of references.

%%%%%%%%%%%%%%%%%%%%%%%%%%%%%%%%%%%%%%%%%%%%%%%%%%

% Don't change these lines
\bsp	% typesetting comment
\label{lastpage}
\end{document}